\documentclass[11pt]{article}
\usepackage{epsfig}
\newcommand{\sss}{\vspace{.2in}}

\newcommand{\be}{\begin{equation}}
\newcommand{\ee}{\end{equation}}
\newcommand{\bea}{\begin{eqnarray}}
\newcommand{\eea}{\end{eqnarray}}
\newcommand{\sn}{{\rm sn}}
\newcommand{\cn}{{\rm cn}}
\newcommand{\dn}{{\rm dn}}
\newcommand{\sech}{{\rm sech}}
\def\Blb{\left[}
\def\Brb{\right]}
\begin{document}
~\hfill{\footnotesize UICHEP-TH/99-3, IP/BBSR/99-10,~~\today}
\sss
\sss
\begin{center}
{\Large {\Large \bf New Solvable and Quasi Exactly Solvable Periodic Potentials}}
\end{center}
\vspace{.5in}
\begin{center}
{\large{\bf
   \mbox{Avinash Khare}$^{a,}$\footnote{khare@iopb.res.in} and
   \mbox{Uday Sukhatme}$^{b,}$\footnote{sukhatme@uic.edu}
 }}
\end{center}
\vspace{.6in}
\noindent
a) \hspace*{.2in}
Institute of Physics, Sachivalaya Marg, Bhubaneswar 751005, Orissa, India\\
b) \hspace*{.2in}
Department of Physics, University of Illinois at Chicago, Chicago, IL 60607-7059, U.S.A. \\
\sss
\sss
\begin{abstract}
Using the formalism of supersymmetric quantum mechanics, we obtain 
a large number of new analytically solvable one-dimensional periodic potentials 
and study their properties. More specifically, the supersymmetric partners 
of the Lam\'{e} potentials $ma(a+1)~{\rm sn}^2(x,m)$ are 
computed for integer values $a=1,2,3,...$. 
For all cases (except $a=1$), we show that the partner potential is distinctly 
different from the original Lam\'{e} potential,  
even though they both have the same energy band structure. 
We also derive and 
discuss the energy band edges of the associated Lam\'{e} potentials 
$pm~{\rm sn}^2(x,m)+qm~{\rm cn}^2(x,m)/{\rm dn}^2(x,m)$, which constitute 
a much richer class of periodic problems. Computation of their 
supersymmetric partners yields many additional new 
solvable and quasi exactly solvable 
periodic potentials.
\end{abstract}
\newpage

\sss
{\noindent\bf 1. Introduction:} 

The energy spectrum of electrons on a lattice is of central importance in 
condensed matter physics. In particular, knowledge of the existence and 
locations of band edges and band gaps determines many physical properties.  Unfortunately,
even in one dimension, there are very few analytically solvable periodic 
potential problems in quantum mechanics. The aim of this paper is to  
extend the small currently known set of analytically solvable periodic potentials. 

For a potential with period $L$, 
one is seeking solutions of the Schr\"{o}dinger equation subject to the 
Bloch condition
\be\label{1}
\psi (x) = e ^{ikL} \ \psi (x+L) \, ,
\ee
where $k$ denotes the crystal momentum. The spectrum shows energy bands whose 
edges correspond to $kL=0,\pi$, that is the wave functions at the band 
edges satisfy $\psi (x) = \pm \psi (x+L)$. For periodic potentials, the 
band edge energies and wave functions are often called eigenvalues and 
eigenfunctions, and we will also use this terminology. The classic text book example 
which is used to demonstrate band structure is the Kronig-Penney model 
$$V(x) = \sum^{\infty}_{n=-\infty} \ V_0 \delta (x-nL) \, ~.$$ 
It should be noted that the band edges for the Kronig-Penney model can only 
be computed by solving a transcendental equation. Another well studied class of 
periodic potentials is
\be\label{3}
V(x) = pm~{\rm sn}^2(x,m) ~,~ p \equiv a(a + 1) ~~.
\ee
Here ${\rm sn}(x,m)$ is a Jacobi elliptic function of real elliptic modulus
parameter $m$ $( 0\leq m\leq 1)$ with period $4K(m)$. For simplicity, from 
now onward, we will not explicitly display the modulus parameter $m$ as an 
argument of Jacobi elliptic functions \cite{gr}. The elliptic 
function potentials of eq. (\ref{3}) have a period $L=2K(m),$ and will be 
referred to as Lam\'{e} potentials, since the 
corresponding Schr\"{o}dinger 
equation is called Lam\'{e}'s equation \cite{ar,mw}.  
It is well known 
that for any integer value $a=1,2,3,\ldots$, the corresponding 
Lam\'{e} potential (\ref{3}) has $a$ bound bands followed by a continuum band 
\cite{ar,mw}. All band edge energies and wave functions are analytically known.

At this point it is worth recalling that supersymmetric 
quantum mechanics (SUSYQM) has proved useful in discovering many, new, 
analytically solvable potentials on both the full as well as the half line 
\cite{CKS}. It is then 
natural to enquire if one can also use similar techniques 
to discover new solvable periodic potentials. In this paper, we demonstrate 
that this is indeed possible. 

Our work is 
inspired by several recent papers \cite{df,dm,bm,bg} which discuss various 
general aspects of SUSYQM for periodic potentials.
In particular, Dunne and Feinberg \cite{df} defined and developed the concept 
of ``self-isospectral" periodic potentials in detail. A one 
dimensional potential $V_-(x)$ of period $L$ is said to 
be self-isospectral if its supersymmetric partner 
potential $V_+(x)$ is just the original potential upto a 
discrete transformation - a translation by any constant amount, a 
reflection, or both. A common example is translation by half a 
period, in which case the condition for self-isospectrality is 
\be\label{8}
V_+(x)=V_-(x-L/2)~~. 
\ee
It is easily checked that if the superpotential $W$ satisfies
\be\label{8a}
W(x) = - W(x-L/2) \, ,
\ee
then condition (\ref{8}) immediately follows.
In this 
sense, any self-isospectral potential is rather uninteresting, since  
application of the SUSYQM formalism \cite{CKS} to it just yields a discrete transformation and basically nothing new. 
We have recently pointed out \cite{sk} that the Lam\'{e} 
potentials given in eq. (\ref{3}) are not self-isospectral for $a \ge 2$, 
and hence SUSYQM generates new exactly solvable periodic problems. This point 
is further developed in detail in this paper. 

We expand our discussion to the band edges and wave functions of a much richer class of periodic potentials given by
\be\label{61}
V(x) = pm~{\rm sn}^2 (x) + qm~{{\rm cn}^2(x)\over {\rm dn}^2 (x)}~,~ ~~p \equiv a(a+1)~,~q \equiv b(b+1)~~,
\ee
where, like ${\rm sn}(x)$, the Jacobi elliptic functions ${\rm cn}(x)$ and 
${\rm dn}(x)$ also have a modulus parameter $m$ which, for 
notational convenience, is not explicitly displayed. The potentials of eq. (\ref{61}) are called associated Lam\'{e} potentials, since the corresponding Schr\"odinger equation is called the associated Lam\'{e} equation \cite{mw}. More precisely, we often refer to the 
associated Lam\'{e} potential of eq. (\ref{61}) as the $(p,q)$ potential and note 
that $(p,0)$ potentials are just the ordinary Lam\'{e} potentials. Although some 
results for $(p,q)$ potentials are available in scattered form in the 
mathematical literature, many of our results are new. In particular, 
we obtain band edge energies and wave functions for the special case 
$p=q=a(a+1)$ for $a=1,2,3,\ldots$. 
We study 
many $(p,q)$ potentials and check whether they are self-isospectral by constructing and examining the supersymmetric partner potentials. In most cases, 
$V_-(x)$ is not self-isospectral, and consequently $V_+(x)$ is a new, 
exactly or quasi-exactly solvable periodic potential.

The associated 
Lam\'{e} potentials given by eq. (\ref{61}) can also be re-written in 
the alternative form
\be \label{a2}
V(x)=pm~\sn^2(x)+qm~\sn^2(x+K(m))~~,~
\ee
since \cite{gr} $$\sn(x+K)=\cn(x)/\dn(x)~,~\cn(x+K)=
-~\sqrt{1-m} ~\sn(x)/\dn(x)~,~\dn(x+K)=\sqrt{1-m}/\dn(x)~.$$ It is clear from (\ref{a2}) that potentials 
$(p,q)$ and $(q,p)$ have the same energy spectra with wave functions shifted by $K(m)$. 
Therefore, it is sufficient to restrict our attention to $p \ge q.$

Before actually solving the Schr\"odinger equation for the associated Lam\'e potential (\ref{61}), let us make a few general comments. Throughout this paper, we have chosen units with $\hbar=1,$ and taken the particle mass in the Schr\"odinger equation to be 1/2. 
Note that in the limit when the elliptic modulus parameter $m=0$, the potential vanishes and one
has a rigid rotator problem of period $2K(0) = \pi,$ whose energy eigenvalues are at $E=0,1,4,9,...$ with all
the nonzero values being two-fold degenerate.
On the other hand, the limit $m \rightarrow 1$ is much trickier since $K(m)$ tends to infinity and the periodic nature of the potential is obscured.  The Schr\"odinger equation for finding the eigenstates for an arbitrary periodic potential is called Hill's equation in the mathematics literature \cite {mw}. A general property of Hill's equation is the oscillation theorem which states that for a potential with period $L$, the band edge wave functions arranged in order of increasing energy $E_0 \le E_1 \le E_2 \le E_3 \le E_4 \le E_5 \le E_6 \le ...$
are of period $L,2L,2L,L,L,2L,2L,...$. The corresponding number of wave function nodes in the interval $L$ are $0,1,1,2,2,3,3,...$ and the energy band gaps are given by $\Delta_1 \equiv E_2 - E_1,~\Delta_2 \equiv E_4 - E_3,~ \Delta_3 \equiv E_6 - E_5,~... $. We shall see that the expected $m=0$ limit and the oscillation theorem are very useful in identifying if all band edge eigenstates have been properly determined or if some have been missed. 

The plan of the paper is as follows. In Sec. 2, we briefly review 
the basic ideas of SUSYQM. A detailed discussion of Lam\'{e} 
potentials and their supersymmetric partners 
is given in Sec. 3. Solutions of the 
Schr\"{o}dinger equation for the associated Lam\'{e} potentials are 
presented in Sec. 4. Many key new results are summarized in Table 3. It is shown that 
the locus of quasi exactly solvable problems \cite{tu,us} 
in the $(p,q)$ plane are parabolas about the line $p=q$. Our solutions are valid for any real choice of the parameters $a,b$ (recall $p=a(a+1),q=b(b+1)$). 
Integer and half-integer values of $a,b$ ,  
including the very interesting special case $a = b =$integer, are 
treated in detail in Sec. 5.  In most cases, the application of SUSYQM gives new solvable periodic potentials, many of which are illustrated in the figures. 
Finally, Sec. 6 contains some concluding remarks. 

\sss
{\noindent\bf 2. Supersymmetric Quantum Mechanics Formalism:} 

The supersymmetric partner potentials $V_{\pm}(x)$
are defined in terms of the superpotential $W(x)$ by
\be\label{9}
V_{\pm} (x) = W^2 (x) \pm W'(x) \, .
\ee
The corresponding Hamiltonians $H_{\pm}$ can be factorized as
\be\label{10}
H_{-} = A^+ A, \ H_{+} = A A^+ \, ,
\ee
where
\be\label{11}
A = {d\over dx} + W(x) \, , ~A^+ = -{d\over dx} + W(x) \, ,
\ee
so that the spectra of $H_{\pm}$ are nonnegative. It is also
clear that
on the full line, both $H_{\pm}$ cannot have zero energy modes since
both $\psi_0^{(\pm)}$ given by
\be\label{12}
\psi_0^{(\pm)} (x) = \exp (\pm \int^x W(y) dy) ~~ ,
\ee
cannot be simultaneously normalized.

On the other hand, when the superpotential
$W(x)$ is
periodic ($W(x+L)=W(x)$) then the potentials $V_- (x)$  and $V_+ (x)$ are isospectral - their spectra match
completely,  including the zero modes, and SUSY is unbroken
provided
\be\label{13}
\int^L_0 W(y) dy = 0 \, .
\ee
It is worth noting that in this case both $\psi_0^{(\pm)}$ belong to the 
Hilbert space. Thus in this
case even though SUSY is unbroken, the Witten index is zero \cite{df}. The condition
(\ref{13}) is trivially satisfied in case $W(x)$ is an odd function of $x$
and throughout this paper we shall only consider superpotentials $W$ which are odd 
function of $x$.
Further, using the
known eigenfunctions $\psi^{(-)}_n (x)$ of $V_{-}(x)$
one can
immediately
write down the corresponding un-normalized eigenfunctions $\psi^{(+)}_n (x)$
of $V_{+} (x)$. In particular, from eq. (\ref{12}) it follows that the 
ground state of $V_{+} (x)$ is given by
\be\label{14}
\psi^{(+)}_{0} (x) = {1\over \psi^{(-)}_{0} (x)} \,~ ,
\ee
while the excited states $\psi^{(+)}_{n} (x)$ are
obtained from $\psi^{(-)}_{n} (x)$
by using the relation 
\be\label{15}
\psi^{(+)}_{n} (x) = \Blb{d\over dx} +W(x)\Brb \psi^{(-)}_{n} (x) ~,~ (n \ge 1)~.
\ee
Thus by starting from an exactly solvable periodic potential $V_-(x)$, 
one gets another isospectral periodic
potential $V_+(x)$. As emphasized previously, if $V_-(x)$ is not self-isospectral, then $V_+(x)$ 
is a new solvable periodic potential! 

\sss
{\noindent\bf 3. Lam\'{e} Potentials (p,0) and Their Supersymmetric Partners:} 

The supersymmetric quantum mechanics formalism of the previous section will now be applied to the Lam\'{e} 
potentials $~ma(a+1)~{\rm sn}^2(x,m).$  
Analytic solutions are known for integer values of $a$ \cite{ar}, and the supersymmetric partner potentials can be readily computed. We first discuss the results for small integer values of $a$, and then present some eigenstate results for arbitrary integer values of $a$. \sss

\noindent{\bf A. Lam\'{e} potentials with a = 1,2,3 :}

{\bf a=1:}	The $a=1$ Lam\'{e} potential $V_- = 2m\, \sn^2(x)-m$ 
is known to be self-isospectral \cite{df} since its SUSY partner satisfies $V_+(x)=V_-(x-K(m))$. 
Both $V_+(x)$ and $V_-(x)$ have one  
energy band ranging from energy $0$ to 
energy $1-m$, with a continuum starting at energy $1$ \cite{ar}. Note that 
at $m=0$ one has energy eigenvalues at $0,1$ as expected for a rigid rotator and as $m \rightarrow 1$, one gets $V_-(x) \rightarrow 1 - 2 \sech^2 x$, 
the band width $1-m$ vanishes as expected, and one has an energy level at $E=0$.

{\bf a=2:}	For the $a$ = 2 case, 
the Lam\'{e} potential (\ref{3}) has 2 bound bands and a
continuum band. The energies and wave functions of the
five band edges are well known \cite{ar,mw}. The lowest energy band 
ranges from $2+2m-2\delta$ to $1+m$, the second energy band ranges 
from $1+4m$ to $4+m$ and the continuum starts at energy $2+2m+2\delta$, 
where $\delta = \sqrt{1-m+m^2}$. The wave functions of all the band edges 
are given in Table 1. Note that in the 
interval $2K(m)$ corresponding to the 
period of the  Lam\'{e} potential, the number of nodes increases with energy.
In order to use the SUSYQM formalism,
we must shift the Lam\'{e} potential by a constant to ensure
that the ground state i.e. (the lower edge of the lowest band) has energy
$E = 0$. As a result, the potential 
\be\label{16}
V_{-} (x) = -2-2m+2\delta+6m {\rm sn}^2(x)
\ee
has its ground state energy at zero with a corresponding un-normalized 
wave function \cite{ar}
\be\label{17}
\psi^{(-)}_{0} (x) = 1+m+\delta-3m {\rm sn}^2 (x)~~.
\ee
The corresponding superpotential is
\be\label{19}
W = - {d\over dx} \log \psi^{(-)}_{0} (x)
= {6m {\rm sn}(x) {\rm cn}(x) {\rm dn}(x)\over
\psi^{(-)}_{0} (x)} \, ,
\ee
and hence the partner potential $V_{+} (x)$ for the potential $V_{-} (x)$ given in eq. 
(\ref{16}) is
\be\label{20}
V_{+}(x) = -V_-(x)+ 
{72 m^2 {\rm sn}^2(x) {\rm cn}^2(x) {\rm dn}^2(x)\over [1+m+\delta-3m {\rm sn}^2 (x)]^2}~~.
\ee
Although the SUSYQM formalism guarantees that the potentials $V_{\pm}$ 
are isospectral, they are not
self-isospectral, since they do not satisfy 
eq. (\ref{8}) \cite{sk}.
Therefore, $V_+(x)$ as given by eq. (\ref{20}) is a new 
periodic potential which is strictly isospectral
to the potential (\ref{16}) and hence it also has 2 bound bands and a 
continuum band. In Fig. 1 we have plotted the
potentials $V_{\pm}(x)$ corresponding to $a = 2$ for three
different values of the parameter $m$. The values are $m=0.5, 0.8,0.998$. 
The difference in 
shape between $V_-(x)$ and $V_+(x)$ is manifest from the figures, especially for large $m$. Using 
eqs. (\ref{14}) and (\ref{15})
and the known eigenstates of $V_{-} (x)$,
we can immediately compute all the band-edge Bloch wave functions for 
$V_{+} (x)$. In Table 1 
we have given the energy eigenvalues and wave functions for the isospectral 
partner potentials $V_{\pm} (x)$. At 
$m=0$ one has energy eigenvalues $0,1,4$ as expected for a rigid rotator. As $m \rightarrow 1$, one gets 
$V_-(x) \rightarrow 4 - 6 \sech^2 x$, 
the band widths vanish as expected, and one has two energy levels at $E=0,3$, with a continuum above $E=4$.

{\bf a=3:}	For the $a = 3$ Lam\'{e} potential, the ground state wave function is 
$$\psi^{(-)}_{0} (x) = {\rm dn}(x) [2m+\delta_1+1 -5m \,{\rm sn}^2(x)] ~~,~ $$
the corresponding superpotential is \cite{sk}
\begin{equation}
W = {m \,{\rm sn}(x) {\rm cn}(x)\over {\rm dn}(x)}
\ {[2m+\delta_1+11-15m\, {\rm sn}^2(x)]\over
[2m+\delta_1+1-5m \,{\rm sn}^2(x)]} ~~,~ 
\end{equation} 
and the partner potentials $V_{\pm} (x)$ are \cite{sk}

\be\label{20c}
V_{-}(x) = -2-5m+2\delta_1+12m ~{\rm sn}^2(x) ~ ,~\delta_1 
\equiv \sqrt{1-m+4m^2} ~ ,
\ee
and
\be\label{20d}
V_{+}(x) = -V_{-}(x) +{2m^2 {\rm sn}^2(x) {\rm cn}^2(x)\over
{\rm dn}^2(x)} {[2m+\delta_1+11-15m ~{\rm sn}^2(x)]^2\over
[2m+\delta_1+1-5m ~ {\rm sn}^2(x)]^2} ~~ .
\end{equation}
Clearly, the
potential $V_{-}(x)$ is not self-isospectral.
In fact, $V_{-}(x)$ and $V_{+}(x)$ are distinctly different periodic 
potentials which have the same seven  
band edges corresponding to three bound bands
and a continuum band \cite{ar}. 
In Fig. 2 we have
plotted the potentials
$V_{\pm}(x)$ corresponding to $a = 3$ for several
different values of the parameter $m$. 
The values of $m$ are 0.5, 0.8, 0.998.
It is clear from the figure that the
potentials $V_{+} (x)$ and $V_{-} (x)$ have different shapes and 
are far from being self-isospectral.
Using eqs. (\ref{14}) and (\ref{15})
and the known eigenstates of $V_{-} (x)$,
we can immediately compute all the 7
band edges corresponding to the known 3 bound bands
and a continuum band \cite{ar,mw}. For
example, the ground state $\psi^{(+)}_{0}$ is given by
\be\label{25}
\psi^{(+)}_{0} (x) = {1 \over \psi^{(-)}_{0} (x)}
= {1\over {\rm dn}(x) [1+2m+\delta_1-5m ~{\rm sn}^2(x)]} ~ .
\ee
The wave functions for the remaining
six states are similarly written down by using eq. (\ref{15}). These are shown in Table 2. The band edge energies for the $a=3$ Lam\'{e} potential (12,0) as a function of the elliptic modulus parameter $m$ are plotted in Fig. 3. Note that 
at $m=0$ one has energy eigenvalues at $0,1,4,9$ as expected for a rigid rotator and as $m \rightarrow 1$, one gets 
$V_-(x) \rightarrow 9 - 12 \sech^2 x$, 
the band widths vanish as expected, and one has three energy levels at $E=0,5,8$ with a continuum above $E=9$.
\sss

\noindent {\bf B. Results for general integer values of a:}

The extension to higher values of $a$ is straightforward.
It is possible to make several general comments about the 
form of the band edge wave functions for the partner potentials $V_{+} (x)$. 
This is most conveniently done by separately discussing the cases of 
even and odd values of $a$. 

{\bf  a= even integer:}	For $a$ even, say $a = 2N$, it is known
\cite{ar} that there are  
$N+1$ solutions of the form $F_N (\sn^2 x)$, and $N$ solutions each
of the three forms
$$\sn x ~\cn x ~F_{N-1} (\sn^2 x)~,~ \sn x ~\dn x~ F_{N-1} (\sn^2 x)~,~ 
\cn x~ \dn x ~F_{N-1} (\sn^2 x)~.$$ 
Here $F_r$ denotes
a polynomial of degree $r$ in its argument. The ground state $\psi_{0}^{-} (x)$ (which is the lower edge of the lowest band) is of the form $F_N (\sn^2 x)$.  
It is easily checked using eq. (\ref{15}) that the corresponding partner 
potential 
$V_{+} (x)$ has $N$ solutions each of the four forms 
$${\dn x~ G_N (\sn^2 x)\over \psi_0^{-} (x)}~,~ 
{\sn x ~G_N (\sn^2 x)\over \psi_0^{-} (x)}~,~ 
{\cn x ~G_N (\sn^2 x)\over \psi_0^{-} (x)}~,~
{\sn x ~\cn x ~\dn x \,G_{N-1} (\sn^2 x)\over \psi_0^{-} (x)}~,$$  
while the ground state  
is given by $\psi_0^{+} (x) = 1/\psi_0^{-} (x)$.

{\bf a= odd integer:}	For $a$ odd,
say $a = 2N+1$, it is known \cite{ar} that the Lam\'e potentials have 
$N+1$ solutions each of the three forms
$$ \sn x~ F_N (\sn^2 x)~,~ \cn x~ F_N (\sn^2 x)~,~ \dn x~ F_N (\sn^2 x)$$
and $N$ solutions of the form
$$\sn x ~\cn x ~\dn x~ F_{N-1} (\sn^2 x).$$ 
The ground state $\psi_{0}^{-} (x)$ is of the form $\dn x ~F_N (\sn^2 x)$. 
We can then easily deduce that the corresponding partner potentials 
$V_{+} (x)$ will have $N+1$ solutions each of the two forms  
$${\sn x ~G_{N+1} (\sn^2 x)\over \psi_0^{-} (x)}~,~ 
{\cn x~ G_{N+1} (\sn^2 x)\over \psi_0^{-} (x)}~,$$
and $N$ solutions each of the two forms
$${\dn x~ G_{N+1} (\sn^2 x)\over \psi_0^{-} (x)}, 
{\sn x ~\cn x ~\dn x~ G_{N} (\sn^2 x)\over \psi_0^{-} (x)},$$  
while as usual, the ground state is given by $\psi_0^{+} (x) = 1/\psi_0^{-} (x)$.

In summary, for integral $a$, 
Lam\'e potentials with $a \ge 2$ are not self isospectral. They have distinct
supersymmetric partner potentials even though both potentials  
have the same $(2a+1)$ band edge eigenvalues.\sss

{\noindent\bf 4. Associated Lam\'{e} Potentials (p,q) and Their Supersymmetric Partners:} 

In contrast to the Lam\'{e} potentials discussed above, there seems to be 
no systematic treatment of associated Lam\'{e} potentials in the 
literature. Therefore, we will first devote some time to discuss 
the properties of associated Lam\'{e} potentials, show that they are quasi 
exactly solvable  and then proceed 
to construct and study their isospectral supersymmetric partner potentials.\sss

\noindent {\bf  A.  Description of associated Lam\'{e} potentials:} 

As mentioned before, we will 
refer to the associated Lam\'{e} potentials given by eq. (\ref{61}) 
or the equivalent form eq. (\ref{a2}) as the $(p,q)$ potential.  
The special cases $p=0$, as well as $q=0$, correspond to 
ordinary Lam\'{e} potentials. 

In general, for any value of $p$ and $q$, the associated Lam\'{e} potentials 
have a period $2K(m)$
since $$\sn(x+2K)=-\sn(x)~,~\cn(x+2K)=-\cn(x)~,~\dn(x+2K)=\dn(x)~.$$ However, 
for the special case $p=q,$ eq. (\ref{a2}) shows that the period is $K(m).$  
From a physical viewpoint, if one thinks of a Lam\'{e} potential $(p,0)$ as 
due to a one-dimensional regular array of atoms with spacing $2K(m),$ 
and ``strength" $p$, then the associated Lam\'{e} potential $(p,q)$ 
results from two alternating types of atoms spaced by $K(m)$ 
with ``strengths" $p$ and $q$ respectively. If the two types of 
atoms are identical [which makes $p=q$], one expects a potential of 
period  $K(m).$

Extrema (defined for this discussion as either local or global maxima and minima) of associated Lam\'{e} potentials are easily found by 
setting $dV(x)/dx = 0.$  This gives $$\sn(x)~\cn(x)~[p~\dn^4(x) - q(1-m)]
=0~~.$$ Extrema occur when (i) $\sn(x)=0,$ that 
is $x=0,~\pm 2K(m),~\pm 4K(m),\ldots;$
(ii) $\cn(x)=0,$ that is $x=~\pm K(m),~\pm 3K(m),\ldots;$ (iii) $\dn^4(x) 
= (1-m)q/p~.$ At the points specified by (i) and (ii), one always has extrema and $V(x)$ has 
values $pm$ and $qm$. In addition, since $\dn^4(x)$ has a minimum value $(1-m)^2$ and 
a maximum value unity \cite{gr}, condition (iii) yields additional extrema provided
 $$(1-m)^2 \le (1-m)q/p \le 1~~.$$ For given fixed values of $q$ and $m$, 
this condition has a solution provided $p$ lies in the critical 
range $$q(1-m) \le p \le q/(1-m)~~.$$  Alternatively, for given fixed values of $p$ and $q$ with $p \ge q$, condition (iii) has a solution provided $m$ is greater than the critical value $1-q/p$. 

The associated Lam\'{e} potentials 
for $q=2, m = 0.5$ and several values of $p$ are plotted in Fig. 4(a). 
In the critical range of $p$ 
values $1 \le p \le 4 ~,$ one expects additional extrema, and these are 
clearly seen in Fig. 4(a). In general the period is $2K(0.5)= 3.708 $ , 
but for $p = q = 2$, the period $K(0.5)$ is evident. Note that as $p$ increases, any given extremum changes character. For example, at $x=0,$ as $p$ increases, one goes from a maximum to a local minimum to an absolute minimum. In Fig. 4(b) we have plotted associated Lam\'{e} potentials for $p=4, q=2$ and several values of $m$. As expected from the above discussion, one always sees extrema at the points specified by conditions (i) and (ii), and additional extrema coming from condition (iii) are evident for $m \ge 1/2$.\sss

\noindent {\bf  B.  Solutions of the associated Lam\'{e} equation - parabolas of solvability:}

The associated Lam\'{e} equation is just the Schr\"odinger equation for 
the potential in eq. (\ref{61}).  
\be\label{61a}
-\frac{d^2\psi}{dx^2} 
+ [pm~{\rm sn}^2 (x) + qm~{{\rm cn}^2(x)\over {\rm dn}^2 (x)}-E]\psi =0  ~~.
\ee
On substituting
\be\label{31}
\psi (x) = [{\rm dn}(x)]^{-b} \ y(x) \, ,
\ee
it is easily shown that $y(x)$,
satisfies the Hermite elliptic equation \cite{mw}
\be\label{32}
 y^{''} (x)
+  2bm {{\rm sn}(x) {\rm cn}(x)\over {\rm dn}(x)} y'(x)
 +  [\lambda - (a+1-b)(a+b) m {\rm sn}^2(x)] y(x) = 0 \, ,
\ee
where
\be\label{33}
p = a(a+1)~,~  q = b(b+1)~,~ E = \lambda + mb^2 \, .
\ee
On further substituting
\be\label{34}
{\rm sn}(x) =  \sin t \, , \ \  y(x) \equiv z (t) \, ,
\ee
one obtains Ince's equation
\be\label{35}
(1-m \sin^2 t) z^{''} (t)+(2b-1) m \sin t \cos t~ z'(t)
 +  [\lambda - (a+1-b) (a+b) m \sin^2 t ] z (t) = 0 \, ,
\ee
which is a well known QES equation \cite{mw}.
In particular, on substituting
\be\label{1a}
\cos t  = u \, , \ \ z(t) \equiv  w(u)
 w(u) =  \sum^{\infty}_{n=0} {u^{n} R_{n}\over n!} \, ,
\ee
it is easily shown that $R_n$ satisfies a three-term recursion
relation.
In particular if $a+b+1 = n$
( $n = 1,2,3,...$) then one obtains $n$ QES solutions. 
Actually $n$ QES solutions
are also obtained in case $b-a = -n (n = 1,2,3,...)$
but since $q$ is unchanged
under $b\rightarrow - b - 1$,
no really new solutions are obtained in this case.
The QES solutions for $n = 1,2,3,4,5$ are given in Table 3. 
In particular, for any given choice of $p = a(a+1),$ Table 3 lists the eigenstates of the associated 
Lam\'{e} equation for various values of $q$. 

For $q=a(a-1)$, there is just one 
eigenstate with energy $ma^2$ and wave function $\psi = dn^a (x)$. Since the wave function 
has period $2K(m)$ and is nodeless, this is clearly the ground state wave function of the $(a(a+1),a(a-1))$ potential for any real choice of the parameter $a$. 
The equations $p=a(a+1)$ and $q=a(a-1)$ are the parametric forms of the equation of the 
parabola $(p-q)^2 = 2(p+q)$, which is plotted in Fig. 5 and labeled $P1$. For any point on the parabola, one knows the ground state wave function and energy $E_0 = ma^2$. The parabola $P1$ includes the points (2,0) and (6,2).

For $q=(a-1)(a-2)$, we see from Table 3 that two eigenstates with energies $1+m(a-1)^2$ and $1+ma^2$ are known. Since they have period $4K(m)$ and just one node in the interval $L=2K(m)$, they must correspond to the first and second band edge energies $E_1$ and $E_2$ of the $(a(a+1),(a-1)(a-2))$ potential. Eliminating $a$ from the equations $p=a(a+1)$ and $q=(a-1)(a-2)$ gives the  
``parabola of solvability" $(p-q)^2 = 8(p+q)-12$, which is plotted in Fig. 5 and labeled $P2$.  This parabola includes the points (2,0) and (6,0) which correspond to Lam\'{e} potentials. Similarly, the parabolas of solvability $Pn~(n=0,1,2,...)$ corresponding to $q=(a-n+1)(a-n)$  in Table 3 are plotted. $n$ eigenstates are known for any point on the parabola of solvability $Pn$. \sss 

\noindent{\bf C.  Supersymmetric partner potentials:}  

It is easily checked from Table 3
that the solution corresponding to $q = a(a-1)$ as well as one of the 
$q = (a-2)(a-3)$ 
solutions are nodeless and correspond to the ground state.
Hence, for these cases, one can obtain
the superpotential and hence the partner potential $V_{+}$ and
enquire if $V_{-}$ is self-isospectral. For example, consider
the case of $p = a(a+1), q = a(a-1)$ in which case $W$ is given by
\be\label{36}
W\equiv - {\psi'_0(x)\over \psi_0(x)}
= am {{\rm sn} (x) {\rm cn} (x)\over {\rm dn} (x)} \, ,
\ee
so that the corresponding partner potentials are
\bea\label{37}
V_{-} & = & (a-1)a m {{\rm cn}^2 (x)\over {\rm dn}^2 (x)}
+ m a (a+1) {\rm sn}^2 (x)
- m a^2 ~~,\nonumber \\
V_{+} & = & a(a+1) m {{\rm cn}^2 (x)\over {\rm dn}^2 (x)}
+ m (a-1)a {\rm sn}^2 (x) - m a^2 \, .
\eea
It is easily seen that these partner potentials satisfy eq. (\ref{8}), are
consequently self-isospectral and  
SUSY gives nothing new in this case. It is amusing to note that the superpotential 
$W$ obtained here
was in fact discussed in ref. \cite{df} (see their eq. (32)).

Let us now consider the SUSY partner potential computed from the ground state for the $p=a(a+1), q= (a-2)(a-3)$
case.  It is given by (see Table 3)
\be\label{38}
\psi_0 (x) = [m(a-1)-1-\delta_1+m(2a-1) {\rm sn}^2 (x)] ({\rm dn} (x))^{a-2}~~,
\ee
where
$\delta_1 = \sqrt{1-m+m^2(a-1)^2}$.
The corresponding superpotential
$W$ turns out to be
\be\label{39}
W = {m(a-2) {\rm sn} (x) {\rm cn} (x)\over {\rm dn} (x)}
- {2m(2a-1) {\rm sn} (x) {\rm cn} (x) {\rm dn} (x)
\over [m(1-a)-1-\delta_1 +m(2a-1){\rm sn}^2 (x)]} \, .
\ee
Hence the corresponding partner potentials are
\be\label{40}
V_{-} (x)  = ma(a+1) {\rm sn}^2 (x)
             +  m(a-3)(a-2){{\rm cn}^2 (x)\over {\rm dn}^2 (x)}
-2-m(a^2-2a+2)+2\delta_1 ~ ,
\ee
\be\label{41}
V_{+}(x) = - V_{-}(x) +2 W^2 (x) ~ .
\ee
It is easily checked that these potentials are not self-isospectral
since they do not satisfy the condition (\ref{8}).
Thus one has discovered
a whole class of new elliptic periodic potentials $V_{+} (x)$
as given by eq. (\ref{41}) for which  
three states are analytically known no matter what $a$ is. 
In particular, the energy
eigenfunctions for $V_+$ of these three states are easily obtained by using
the corresponding energy eigenstates of $V_{-}$ as given in Table 3 and using eqs. (\ref{14}) and (\ref{15}). \sss

{\noindent\bf 5. Associated Lam\'{e} Potentials with Special Values of p and q:} 

We shall now discuss associated Lam\'{e} potentials $(a(a+1),b(b+1))$, where $a$ and
$b$ are either both positive integers or half-integers. In most cases, we show that 
although several band edge energies are exactly known from Table 3, one usually does not know 
all the band edge energies, that is one has a quasi exactly solvable problem. However, in the special case of $p=q (a=b=$ integer), we show that all the band edge eigenstates can be obtained and one has an exactly solvable periodic problem.\sss

\noindent {\bf  A.  a,b = integer, a $\ne$ b: }

First, let us note that the Lam\'{e} potentials $(a(a+1),0)$ are in this category when $a=$ integer and $b=0$. For example, when $a=3$, one has the (12,0) potential. We see from Fig. 5 that two parabolas of solvability pass through the point (12,0). From Table 3 it follows that 3 band edges of period $2K(m)$ are obtained from $q=(a-2)(a-3)$ and 4 band edges of period $4K(m)$ are obtained from $q=(a-3)(a-4)$. Altogether, arranging in order of increasing nodes, one has 7 band edges with periods $2K,4K,4K,2K,2K,4K,4K$ with $0,1,1,2,2,3,3$ nodes respectively. There are no missing states, and as discussed in Sec. 3A, this gives three bound bands and a continuum band.

As a second example with $q \ne 0$, consider the (6,2) associated Lam\'{e} potential, that is $p=6,q=2$. In this case, 
taking $a=2,$ one can get five band edges from Table 3 -
one solution of period $2K$ is obtained from $q=a(a-1),$ 
while the remaining four solutions of period $4K$ are obtained from $q=(a-3)(a-4).$
The eigenvalues and eigenfunctions are given in Table 4 along with the number of nodes in one period $2K$. It is clear that there are two solutions of period $2K$ with 2 nodes in the interval $2K$ which have to be present but have not been obtained. This is also clear from the 
$m=0$ limit, since the energies from Table 5 are 0,1,1,9,9 and the states at 4,4 are missing.  Thus, this is
a QES problem. Fig. 6 illustrates the (6,2) associated Lam\'{e} potential and its supersymmetric partner 
for three choices of $m$. The self-isospectral nature of the (6,2) potential is evident from Fig. 6 - it also follows from eqs. (\ref{37}) with $a=2$.
The band edge energies for the (6,2) associated Lam\'{e} potential as a function of the elliptic modulus parameter $m$ is shown in Fig. 7. The two unobtained band edges of period $2K$ will have energies $E=4$ at $m=0$ and $E=3$ at $m=1$.

Let us now discuss the general associated Lam\'{e} potential $(a(a+1),b(b+1))$. Without any loss of generality let
us assume that $a > b$. Using Table 3, we obtain
$(a-b)$  states of period $2K (4K)$ for 
$q=[a-(a-b)][a-(a-b-1)]$ for $(a-b)$ odd (even), 
and $(a+b+1)$ states of period $4K (2K)$for
$q=[a-(a+b+1)][a-(a+b)]$ for $(a-b)$ odd (even). It can be established that some states are missing by looking at the node structure as well as the $m=0$ limit. Hence we again have a QES problem.\sss 

\noindent {\bf  B. a = b = integer: } 

Let us now discuss 
the special case of $p=q=a(a+1),~a=1,2,...$ . 
In this case 
the associated Lam\'{e} potential (\ref{61}) has period $K$, rather than $2K$.  
It then follows from the oscillation theorem that with increasing energy, the band
edges must have periods $K,2K,2K,K,K,...$ and in the $m=0$ limit the eigenvalues
must go to $E=0,4,16,36,...$ with all nonzero eigenvalues being doubly
degenerate. 
It is easy to check from Table 5 
that one case for which we already have exact results is when $p=q=2$. 
In particular, consider the special case $a = 1,$ for which
$V_{-}(x)$ of eq. (\ref{40}) takes the form
\be\label{42}
V_{-}(x) = 2m {\rm sn}^2 (x) + 2m {{\rm cn}^2 (x)\over {\rm dn}^2 (x)}
- 2-m+2\sqrt{1-m} \, .
\ee
Using Table 5,
we can calculate three energy eigenvalues and 
eigenfunctions of $V_{-}$ taking $a=1$ in $q=(a-2)(a-3).$  
These are given in Table 5. 
Whereas the ground state is of period $K$, the 
next two states in Table 5 indeed have period $2K$.  
Using $a = 1$ in eqs. (\ref{38}) to (\ref{41}), 
we find that the corresponding SUSY partner potential is
\be\label{43}
V_{+} (x) = 2-m-2\sqrt{1-m}
-{8\sqrt{1-m} m^2 {\rm sn}^2 (x) {\rm cn}^2(x)
\over [{\rm dn}^2 (x) +\sqrt{1-m}]^2} \, .
\ee
Are the potentials $V_{\pm} (x)$ self-isospectral? Using the relations
\be\label{87}
{\rm sn} (x+K(m)/2) = (1+\sqrt{1-m})^{1/2} \bigg [{\sqrt{1-m} ~{\rm sn} (x)
+{\rm cn} (x) {\rm dn} (x) \over {\rm dn}^2 (x) + \sqrt{1-m}} \bigg ] \, ,
\ee
\be\label{87a}
{\rm cn} (x+K(m)/2) = (1+\sqrt{1-m})^{1/2} (1-m)^{1/4} 
\bigg [{(1+\sqrt{1-m})^{1/2} {\rm cn} (x)
-{\rm sn} (x) {\rm dn} (x) \over {\rm dn}^2 (x) + \sqrt{1-m}} \bigg ] \, ,
\ee
\be\label{87b}
{\rm dn} (x+K(m)/2) = (1-m)^{1/4} 
\bigg [{(1+\sqrt{1-m}) {\rm dn} (x)
-m {\rm sn} (x) {\rm cn} (x) \over {\rm dn}^2 (x) + \sqrt{1-m}} \bigg ] \, ,
\ee
a little algebra reveals that indeed $V_{\pm}$ are self-isospectral and
satisfy eq. (\ref{8}). 

Are the higher members of the $p=q$ family (i.e. $p=q=6,12,20,...$) also 
self-isospectral? If our experience with the Lam\'{e} case is any guide then we
would doubt it. Indeed, we will now show that the (6,6) associated Lam\'{e} potential is not self-isospectral. We get five band edges analytically from Table 3.
In particular, take $a=2$ and consider the case of $q=(a-4)(a-5),$ for
which we know two eigenstates as given in Table 3. In fact, in this case
three more eigenstates can be analytically obtained but the corresponding
eigenvalues and eigenfunctions have not been given in Table 3 since the
energy eigenvalues are solutions of a cubic equation whose exact solution
for arbitrary $a$ can not be written in a compact form. 
However, for $a=2,$ 
we are able to solve the cubic equation and
obtain the three eigenvalues in a closed simple form. In particular consider 
an ansatz of the form
\be\label{45a}
y = A +B {\rm sn}^2 x +D {\rm sn}^4 x \, .
\ee
On substituting this ansatz in eq. (\ref{32}) it is easy to show that the
energy eigenvalue $\lambda (= E -m(a-4)^2$) must obey the cubic equation
\be\label{45b}
\lambda^3 +[28m -20 -12am] \lambda^2 + [64-304m+160ma+32m^2(a-2)(a-3)]
\lambda -64m(2a-3)(2-2m+ma) = 0 \, .
\ee
The solution of this equation is in general quite lengthy but in the special
case of $a=2$ this cubic equation is easily solved 
yielding three eigenvalues in a compact form. On combining them with  
the two levels given in Table 3, we obtain the eigenvalues and 
eigenfunctions of all the five band edges 
for the case $p=q=6$. These are given in Table 6. We have also
verified that these five eigenstates in ascending order of energy indeed 
have periods $K, 2K, 2K, K, K$ respectively and taht the energy eigenvalues have
expected limits at $m=0$. In particular
the associated Lam\'{e} potential $V_{-} (x)$ is 
\be\label{46}
V_{-} (x) = 6m {\rm sn}^2 (x) + 6m {{\rm cn}^2 (x) \over {\rm dn}^2 (x)}
-8 -2m +2 \delta_8 \, , 
\ee
whose ground state energy is zero while the corresponding eigenfunction 
$\psi_{0}^{-}$ is
\be\label{47}
\psi_{0}^{-} (x) = \frac{
\bigg [1-(4-m-\delta_8) {\rm sn}^2 (x) + 
(4-2m-\delta_8) {\rm sn}^4 (x) \bigg ]}{{\rm dn}^2 (x)} \, ,~~\delta_8 = \sqrt{16-16m+m^2} \, .
\ee

Hence the corresponding superpotential is
\be\label{49}
W (x) = \frac{-2m {\rm sn} (x) {\rm cn} (x)}{{\rm dn} (x)} +
\frac{2{\rm sn} (x) {\rm cn} (x)}{{\rm dn} (x) \psi^{-}_{0} (x)} \bigg [
(4-m-\delta_8) -2(4-2m-\delta_8) {\rm sn}^2 (x) \bigg ] \, ,
\ee
and the partner potential $V_{+} (x)$ which is isospectral to $V_{-} (x)$ is
\be\label{70}
V_{+} (x) = -V_{-} (x) + 2W^2 (x) \, .
\ee 
It is not difficult to see that the $W$ as given by eq. (\ref{49}) does not
satisfy the self-isospectral condition (\ref{8a}) and hence unlike the $p=q
=2$ case, the $p=q=6$ potential is {\it not} self-isospectral. In Fig. 8, we
have plotted the potentials $V_{\pm} (x)$ corresponding to $p=q=6$ for 
several different values of the parameter $m$. The figures confirm that the potentials are far from being self-isospectral. Thus we have
obtained a new exactly solvable periodic potential (\ref{70}) 
which has two bound bands and a continuum band, with five band edges and
the corresponding eigenfunctions being exactly known using Table 6 and 
eqs. (\ref{14}) and
(\ref{15}). In Fig. 9, we plot the band edge energies for the (6,6) potential as a function of the elliptic modulus parameter $m$. 

It is also
clear from here that even the higher associated Lam\'{e} potentials with $p=q=
12,20,...$ which have 7,9,... band edges are also exactly solvable in 
principle  and none of them will be self-isospectral, so that in each case
one obtains a new exactly solvable periodic potential. In particular, 
for $p=q=n(n+1)$ there will be $(2n+1)$ band edges in both 
$V_{\pm} (x)$ whose energy eigenvalues can be obtained from Table 3 when
$q$ has the form $[n-2n][n-(2n+1)]$. 
Out of the $(2n+1)$ band edges in $V_{-} (x)$, $(n+1)$
solutions (including the ground state) 
have the form ${F_n (\sn^2 x) \over \dn^n x}$ while $n$
solutions have the form $F_{n-1} (\sn^2 x){\sn x \cn x \over dn^n x}$.
On the other hand, as far as the $(2n+1)$ solutions of the 
partner potential $V_{+}$ are 
concerned, there are $n$ states each of the two forms

$$ {\sn x \cn x G_n (\sn^2 x) \over \dn^{2n-1} x \psi_0^{-} (x)},
{G_{n+1} (\sn^2 x) \over \dn^{2n-1} x \psi_0^{-} (x)},$$
while the ground state (i.e. the lower edge of the lowest band) is given by
$\psi_0^{+} (x) = 1/\psi_0^{-} (x)$. \sss

\noindent{\bf C. a,b = half-integer:}

Let us now specialize to the case when both $a,b$ are half integral with
$a>b$. As an illustration,
let us first consider the case of $a=3/2,b=1/2$ so that $p=15/4,q=3/4$. In
this case, the oscillation theorem requires 
band edges with periods $2K,4K,4K,2K,2K, ...$. 
Using Table 3 and Fig. 5, we see one gets three eigenstates when $q=
(a-2)(a-3)$ with $a=3/2$,  all with
period $2K$. The ground state is at 
$E_0 = {9m\over 4}$ while there are two degenerate levels at $E_3=E_4=4+{m\over 4}$. To understand this degeneracy better, let us go along the parabola of solvability $P2$ given by $q=
(a-2)(a-3)$. The band gap is given by $\Delta_2 \equiv |-2+m+2\sqrt{1-m+m^2(a-1)^2}|$ and is plotted in Fig. 10. It vanishes at $a=3/2$ (15/4, 3/4) potential, and has the correct values $\Delta_2=2 \sqrt{1-m+m^2}-2+m$ for $a=2$ and $\Delta_2=2 \sqrt{1-m+4 m^2}-2+m$ for $a=3$ which correspond to the (6,0) and (12,0) Lam\'{e} potentials. The vanishing of $\Delta_2$ at $a=3/2$ occurs because the eigenfunctions corresponding to $E_3$ and $E_4$ cross over as one goes along the parabola $P2$.

These arguments are easily generalised in case $p=(n+1/2)(n+3/2), q=
(k+1/2)(k+3/2)$ with $n >k$. The energy eigenvalues of $(n-k)$
states can be obtained by using Table 3 in case $q$ is of the form
$q=[n+1/2-(n-k)][n+1/2-(n-k-1)]$ and the corresponding eigenstates
have period $2K (4K)$ depending on whether $(n-k)$ is odd (even). 
On the other hand, the energy of $(n+k+2)$ states is obtained when
$q$ is of the form $q=[n+1/2-(n+k+2)][n+1/2-(n+k+1)]$ and these states
have the same period $2K (4K)$ as the $n-k$ states when $n-k$ is odd
(even). It turns out that the $n-k$ solutions are
in fact common in both and so we only obtain the energy of the $n+k+2$ band edges and all of
them have the same period $2K (4K)$ depending on if $n-k$ is odd (even)
so that it is only a QES problem and not an exactly solvable problem
as one is unable to obtain a single eigenstate with period $4K (2K)$
in case $n-k$ is odd (even).

We would like to point out some of the pecularities of the spectrum in
these cases. For example, in case 
$(p,q) = (35/4,3/4), (63/4,3/4), (99/4,3/4)...$  then one finds that 
3,4,5,... QES energy levels of period $4K, 2K, 4K,...$ respectively are 
analytically known of which
the one at the highest energy is doubly degenerate.   
As an illustration, in
Table 7 we have given the 4 QES energy eigenstates all of period $2K$ for the 
$(63/4,3/4)$ potential. The interesting point about this case is that the partner potentials $V_{\pm} (x)$ are not self-isospectral and hence
one has discovered a new QES potential where 4 band edges of period
$2K$ and the corresponding eigenfunctions are explicitly known. Of these, the
one at $E = 16+{m\over 4}$ is doubly degenerate, again due to crossover of energy levels.    
Using the ground state wave function, the superpotential is computed to be 
\be
W = {3m\over 2} {{\rm sn} (x) {\rm cn} (x) \over {\rm dn} (x)}
-{24m {\rm sn}(x) {\rm cn}(x) {\rm dn}(x) \over 
[12m {\rm sn}^2 (x) -2 -5m - \sqrt{4-4m+25m^2}]}~~.
\ee
Using eqs. (\ref{14}) and 
(\ref{15}) the eigenstates of the SUSY partner potential $V_{+}$ are then determined.\sss 
 
{\noindent \bf 6. Comments and Conclusions:} 

In this paper, we have discussed solutions of the type given in Table 3, 
which correspond to the parabolas of solvability shown in Fig. 5.  
Lam\'{e} potentials $(p,0)$ with $p = a(a+1)$ and integer $a$, always have two parabolas of solvability passing through - one parabola gives all states of period $2K$ and the other gives all states of period $4K$. This provides a deeper understanding of why such Lam\'{e} potentials are fully solvable \cite{ff}. Similarly, we have obtained eigenstates for a large class of associated Lam\'{e} potentials $(p,q)$. Further, using the formalism of supersymmetric quantum mechanics,
we have been able to discover many new
exactly solvable and quasi exactly solvable periodic potentials involving 
Jacobi elliptic functions. This is a very substantial improvement over the 
currently known small number of exactly solvable periodic problems. 

\sss
{\noindent\bf Acknowledgements:} One of us (US) acknowledges the hospitality 
of the Institute of Physics, Bhubaneswar and partial financial support from 
the U.S. Department of Energy. Both authors thank the Mehta Research 
Institute, Allahabad for hospitality.

\newpage
\noindent{\Large \bf Table Captions}
\vskip .5 true cm

Table 1: The eigenvalues and eigenfunctions for the 5 band edges corresponding
to the $a = 2$ Lam\'{e} potential $V_{-}$ which gives $(p,q)=(6,0)$ 
and its SUSY partner $V_{+}$. Here $B\equiv
1+m+\delta$ and $\delta\equiv \sqrt{1-m+m^2}$. The potentials $V_\pm$ have period $L=2K(m)$ and their analytic forms are given by eqs. (\ref{16}) and (\ref{20}) respectively.  The periods of various eigenfunctions and the number of nodes in the interval $L$ are tabulated. \sss

Table 2: The eigenvalues and eigenfunctions for the 7 band edges corresponding
to the $a = 3$ Lam\'{e} potential $V_{-}$ which gives $(p,q)=(12,0)$
and its SUSY partner $V_{+}$. Here $\delta_1 \equiv \sqrt{1-m+4m^2}; 
~\delta_2 \equiv \sqrt{4-m+m^2}; ~\delta_3 \equiv \sqrt{4-7m+4m^2}$. The potentials $V_\pm$ have period $L=2K(m)$ and their analytic forms are given by eqs. (\ref{20c}) and (\ref{20d}) respectively.  The periods of various eigenfunctions and the number of nodes in the interval $L$ are tabulated. \sss
 
Table 3: Eigenvalues and
eigenfunctions for various associated Lam\'{e} potentials $(p,q)$
with $p = a(a+1)$ and $q = (a-n+1)(a-n)$ for $n = 1,2,3,...$. The periods of various eigenfunctions and the number of nodes in the interval $2K(m)$ are tabulated. Here $~~\delta_4 \equiv \sqrt{1-m+m^2(a-1)^2}~~$;$~~ 
\delta_5 \equiv
\sqrt{4-7m+2ma+m^2(a-2)^2}~~$;$~~\delta_6 \equiv \sqrt{4-m-2ma+m^2(a-1)^2}~~$;
$~~\delta_7 \equiv \sqrt{9-9m+m^2(a-2)^2}~.$ \sss

Table 4: The five eigenvalues and
eigenfunctions for the self-isospectral associated Lam\'{e} potential corresponding to $a=2, b=1$ which gives
$(p,q) = (6,2)$. The potential is 
$V_{-} (x)= 6m {\rm sn}^2 (x) +2m {{\rm cn}^2 (x)
\over {\rm dn}^2 (x)} -4m$, and has period $2K(m).$ The number of nodes in the interval $2K(m)$ is tabulated. \sss

Table 5: The three eigenvalues and
eigenfunctions for the associated Lam\'{e} potential corresponding to $a=b=1$ which gives
$(p,q) = (2,2)$. The potential has period $K(m)$ and the number of nodes in the interval $K(m)$ is tabulated. \sss

Table 6: The five eigenvalues and
eigenfunctions for the associated Lam\'{e} potential corresponding to $a=b=2$ which gives
$(p,q) = (6,6)$. Here $\delta_8 \equiv \sqrt{16-16m+m^2}$ \, . The number of nodes in one period $K(m)$ of the potential  is tabulated. \sss

Table 7: Energy eigenvalues and eigenfunctions for the 
associated Lam\'{e} potential corresponding to $a=7/2, b=1/2$ which gives
$(p,q) = (63/4,3/4)$. Here $\delta_9 \equiv \sqrt{4-4m+25m^2}~$;
$~V_{-} (x)= {63\over 4}m {\rm sn}^2 (x) +{3\over 4}m {{\rm cn}^2 (x)
\over {\rm dn}^2 (x)} -2-{29m \over 4} +\delta_9.$ The last column gives the number of eigenfunction nodes in one period $2K(m)$ of the potential.\sss

\vskip 2.5 true cm

\noindent{\Large \bf Figure Captions}
\vskip .5 true cm
Fig. 1: The (6,0) Lam\'{e} potential $V_-(x)$ corresponding to $a = 2$ [thick line] as given by eq. (\ref{16}) and its supersymmetric partner potential $V_{+} (x)$ [thin line] 
as given by eq. (\ref{20}) for three choices of $m$ (a) 0.5 
(b) 0.8 (c) 0.998. \sss

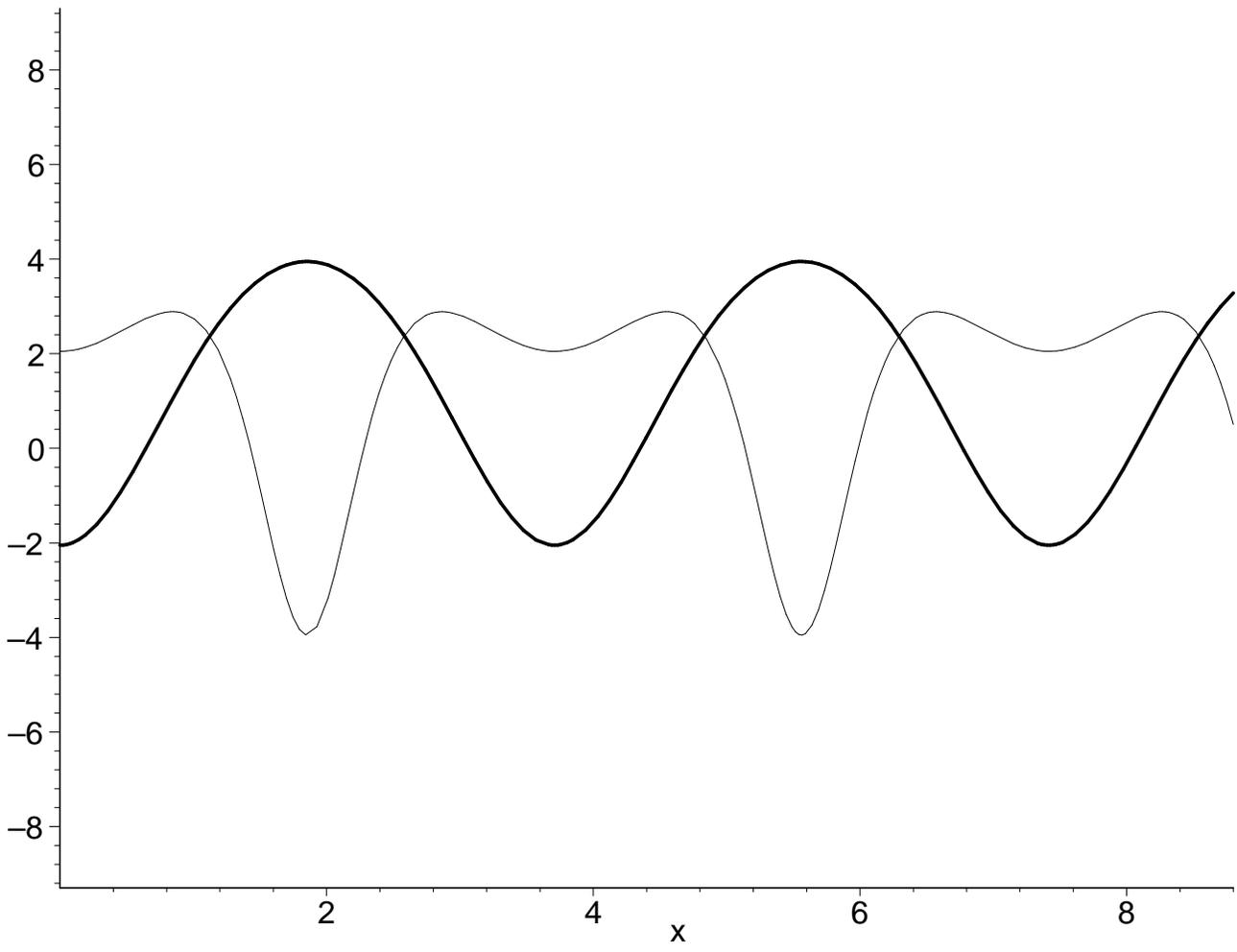
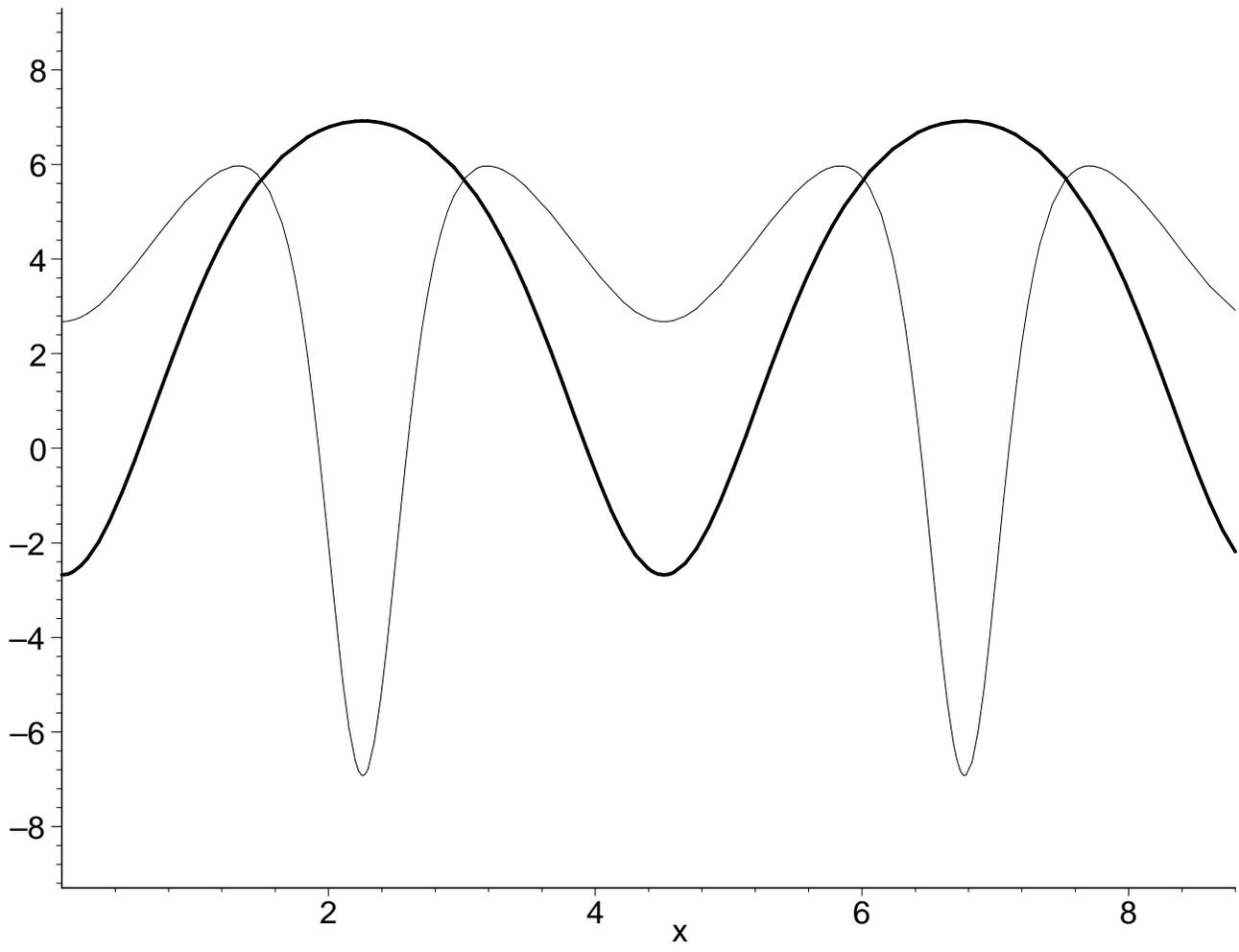
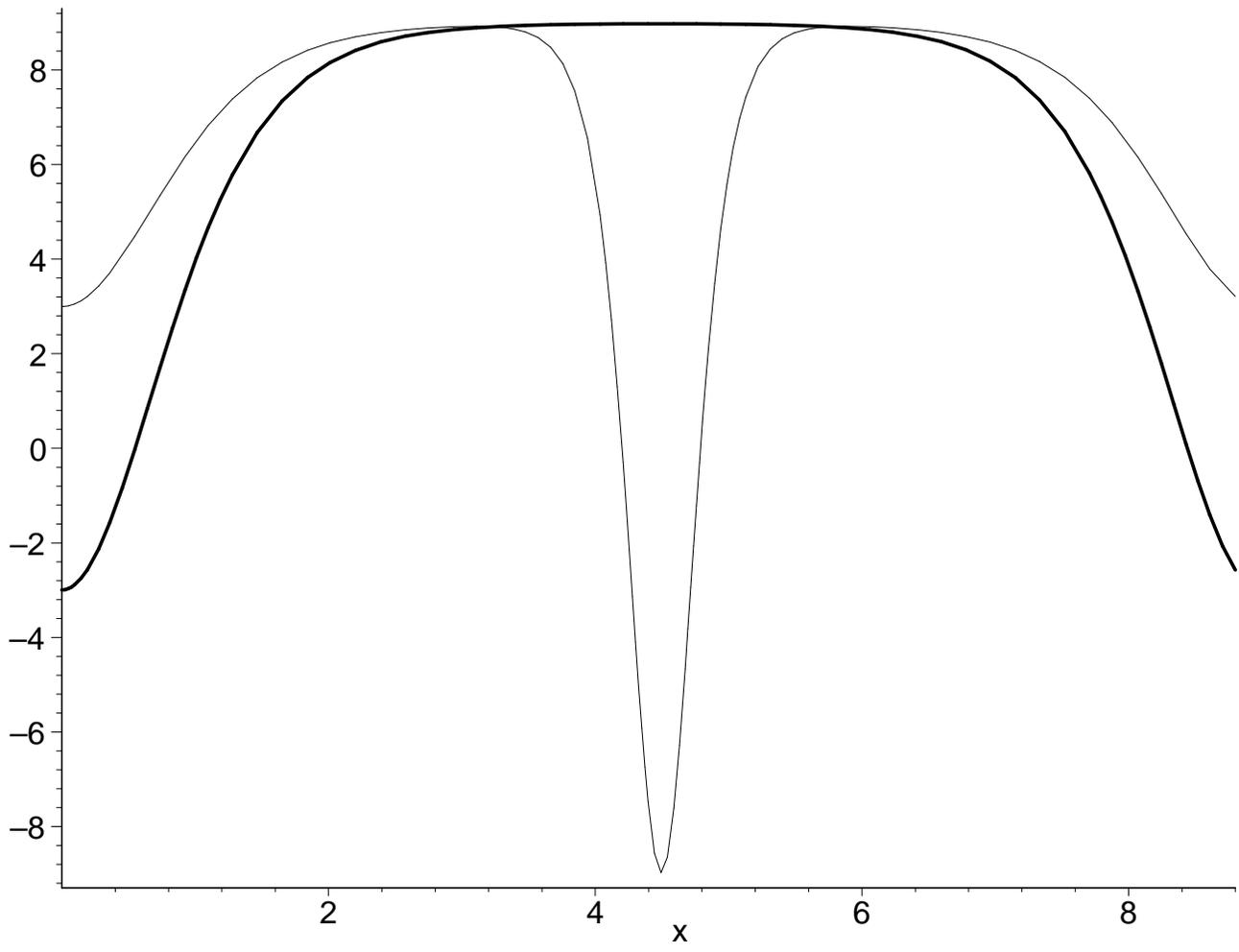
Fig. 2: The (12,0) Lam\'{e} potential $V_-(x)$ corresponding to $a = 3$ [thick line] as given by eq. (\ref{20c}) and its supersymmetric partner potential $V_{+} (x)$ [thin line] 
as given by eq. (\ref{20d}) for three choices of $m$ (a) 0.5 
(b) 0.8 (c) 0.998. \sss

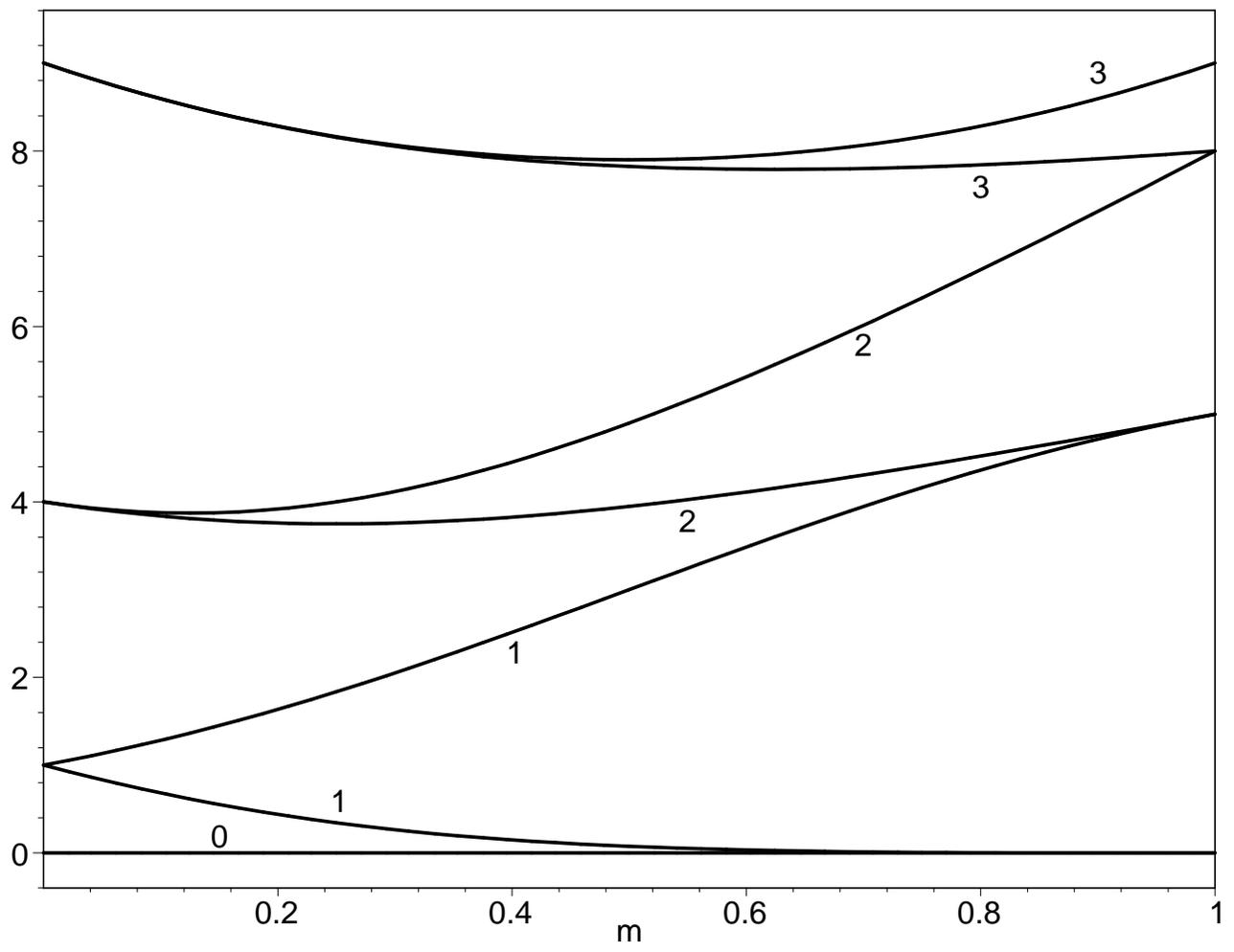
Fig. 3: Band edge energies for the (12,0) Lam\'{e} potential corresponding to $a = 3$ as a function of the elliptic modulus parameter $m$. This figure is drawn using the eigenvalues given in Table 2. \sss

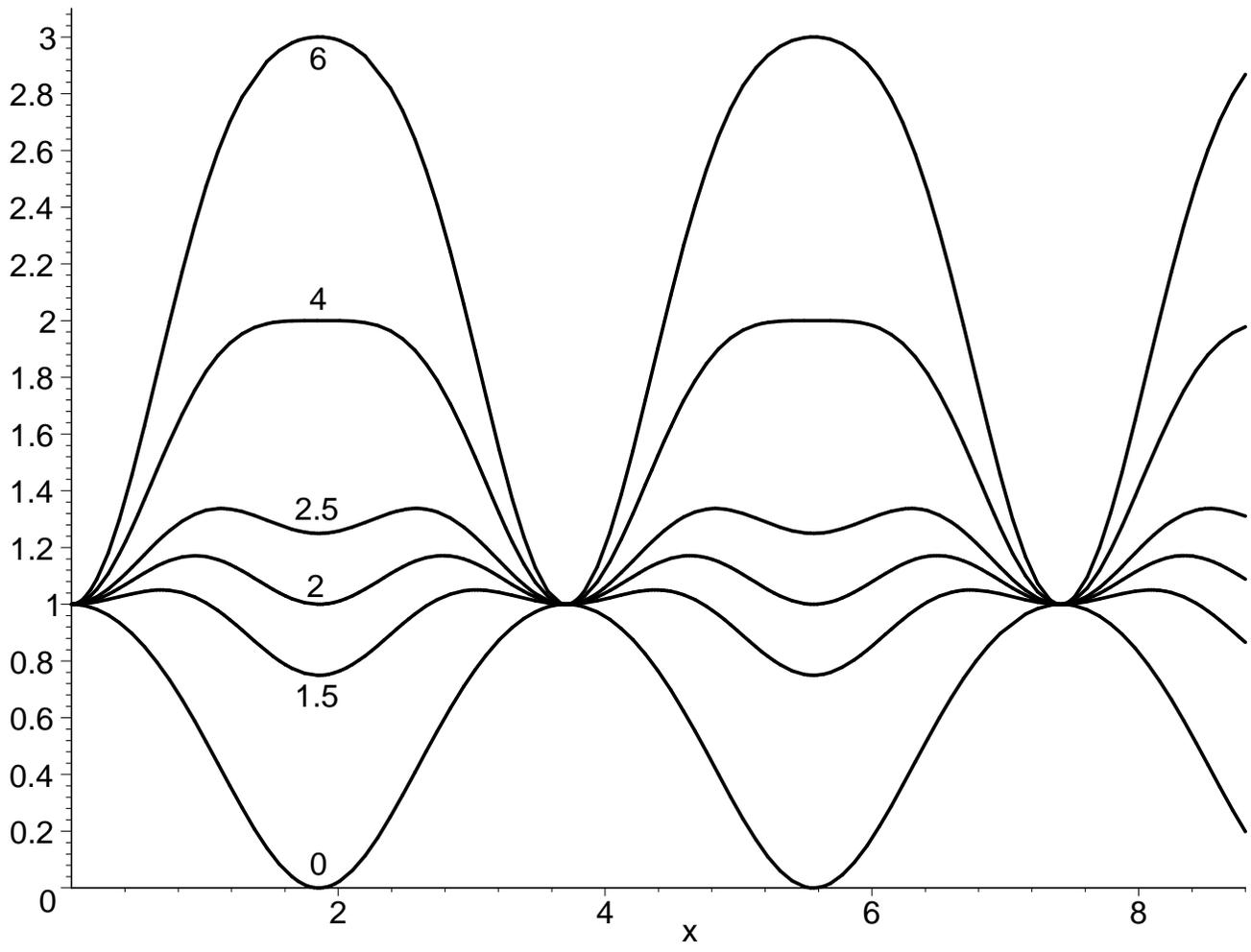
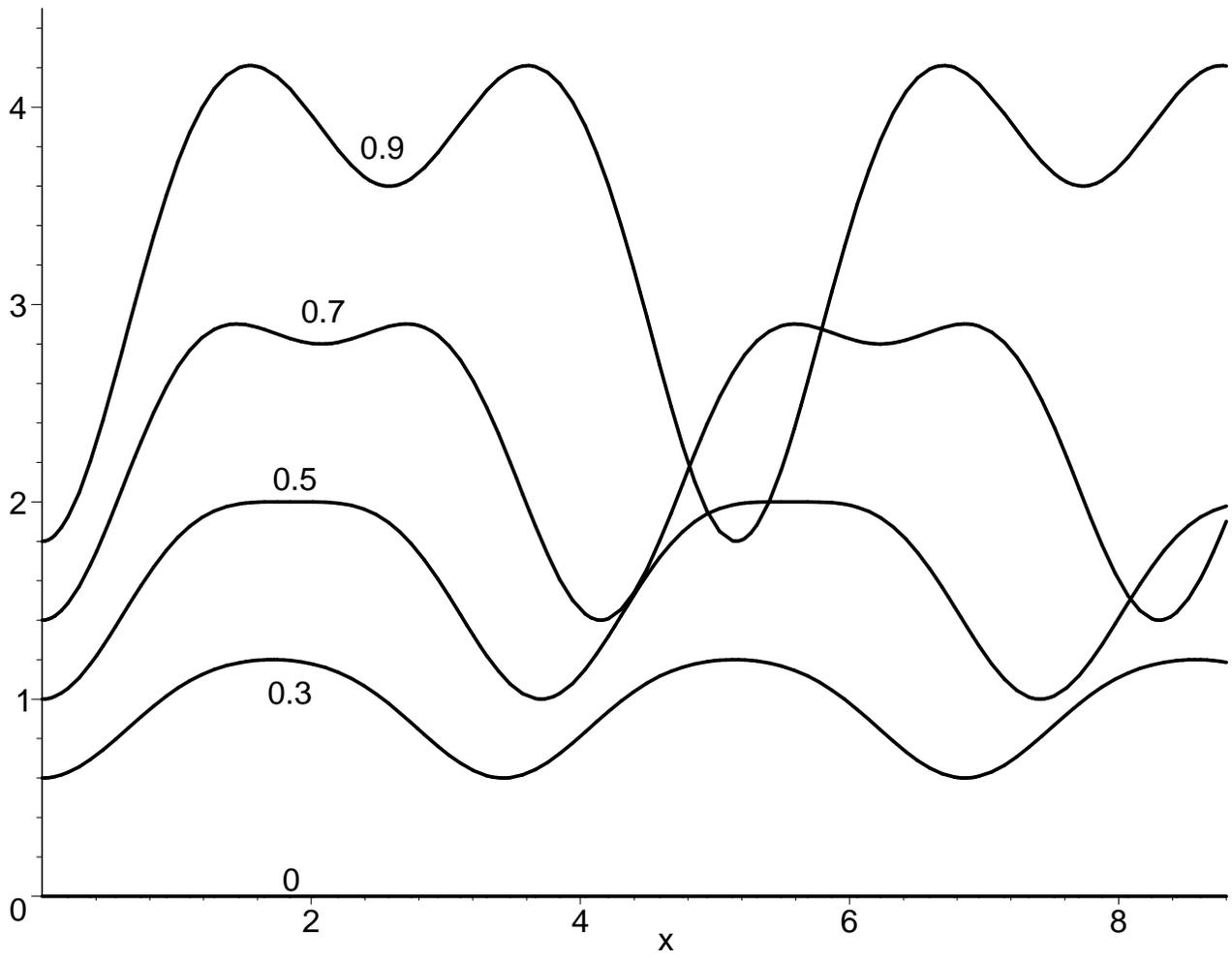
Fig. 4: (a) Plots of the $(p,q)$ associated Lam\'{e} potentials for $q=2, m = 0.5$ and several values of $p$.  (b) Plots of the $(p,q)$ associated Lam\'{e} potentials for $p=4, q = 2$ and several values of $m$.\sss

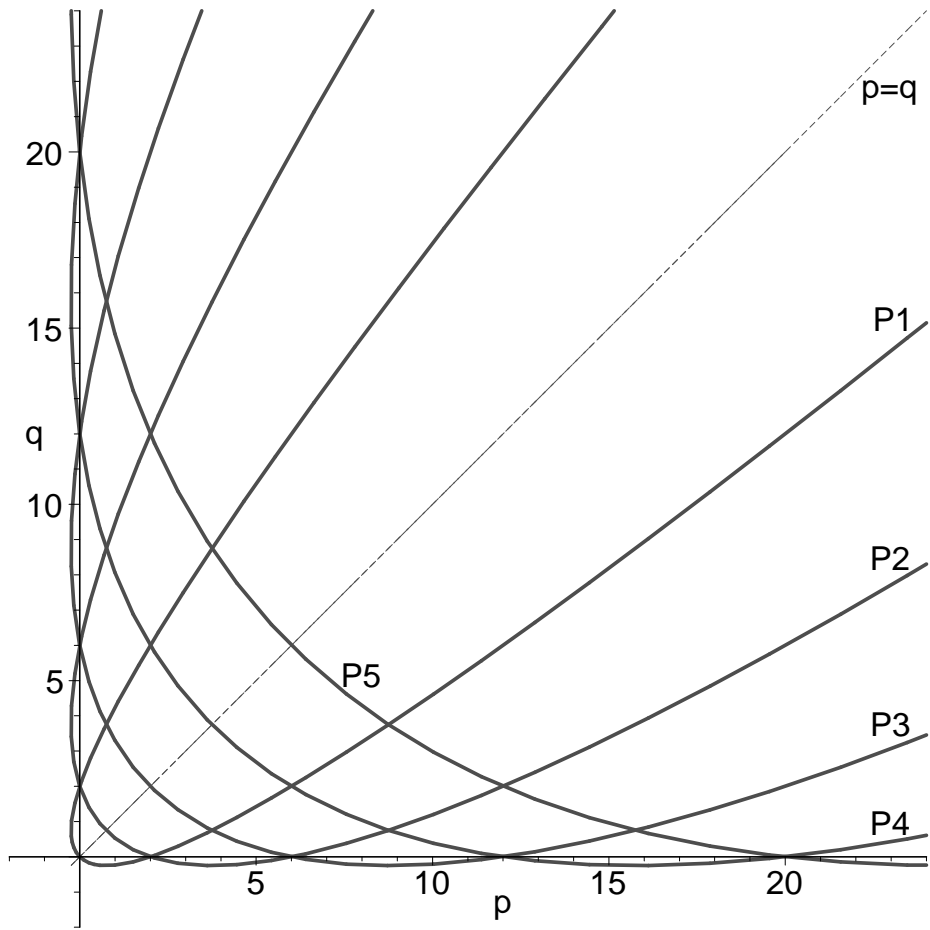
Fig. 5: Parabolas of solvability.  This figure illustrates all associated Lam\'{e} potentials $(p,q)$ which are quasi solvable. Each parabola corresponds to a choice of $q$ in Table 3. Parabola $Pn$ is for $q=(a-n+1)(a-n)$ for $n=1,2,3,...$, and one knows $n$ eigenstates for any point on it from Table 3. \sss

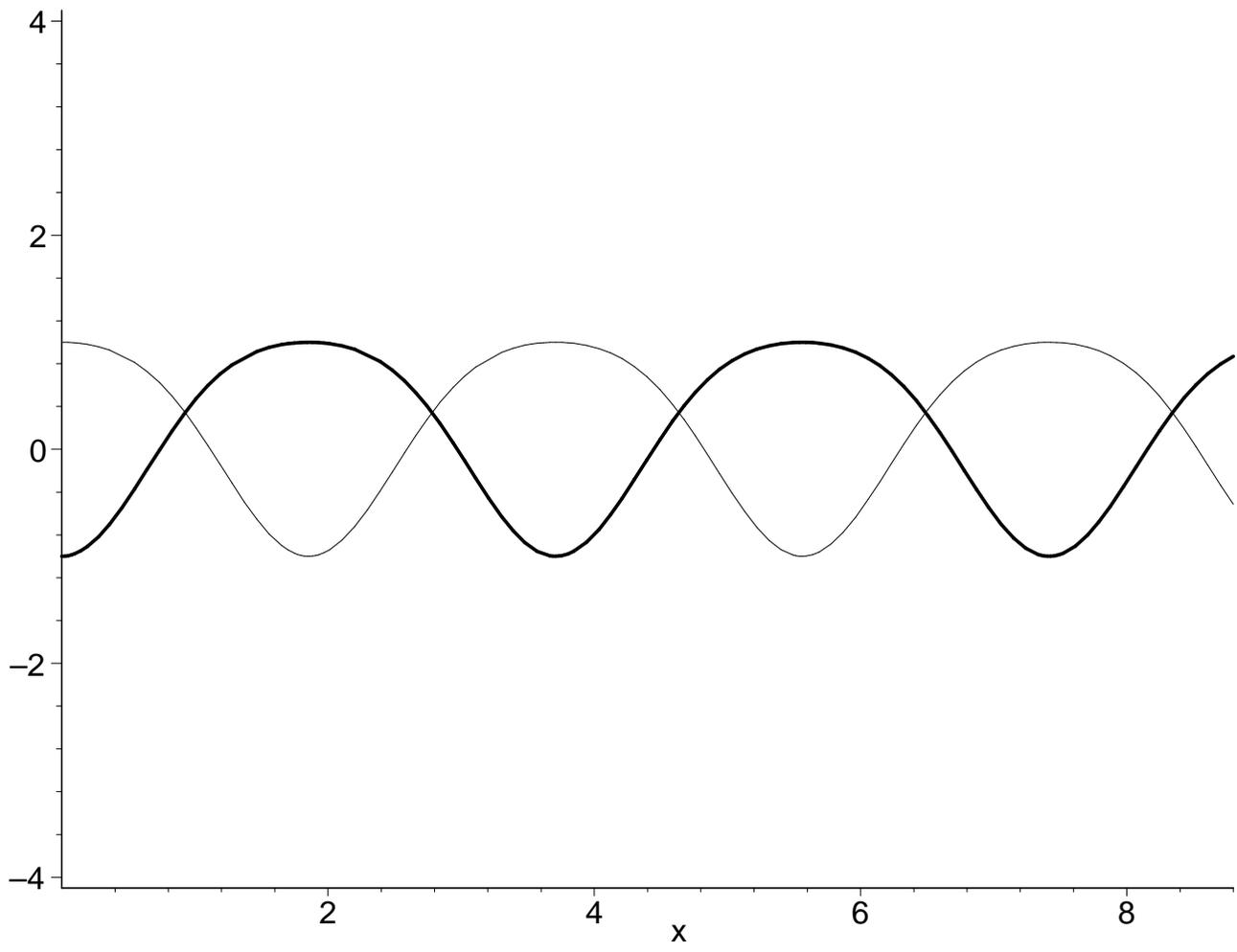
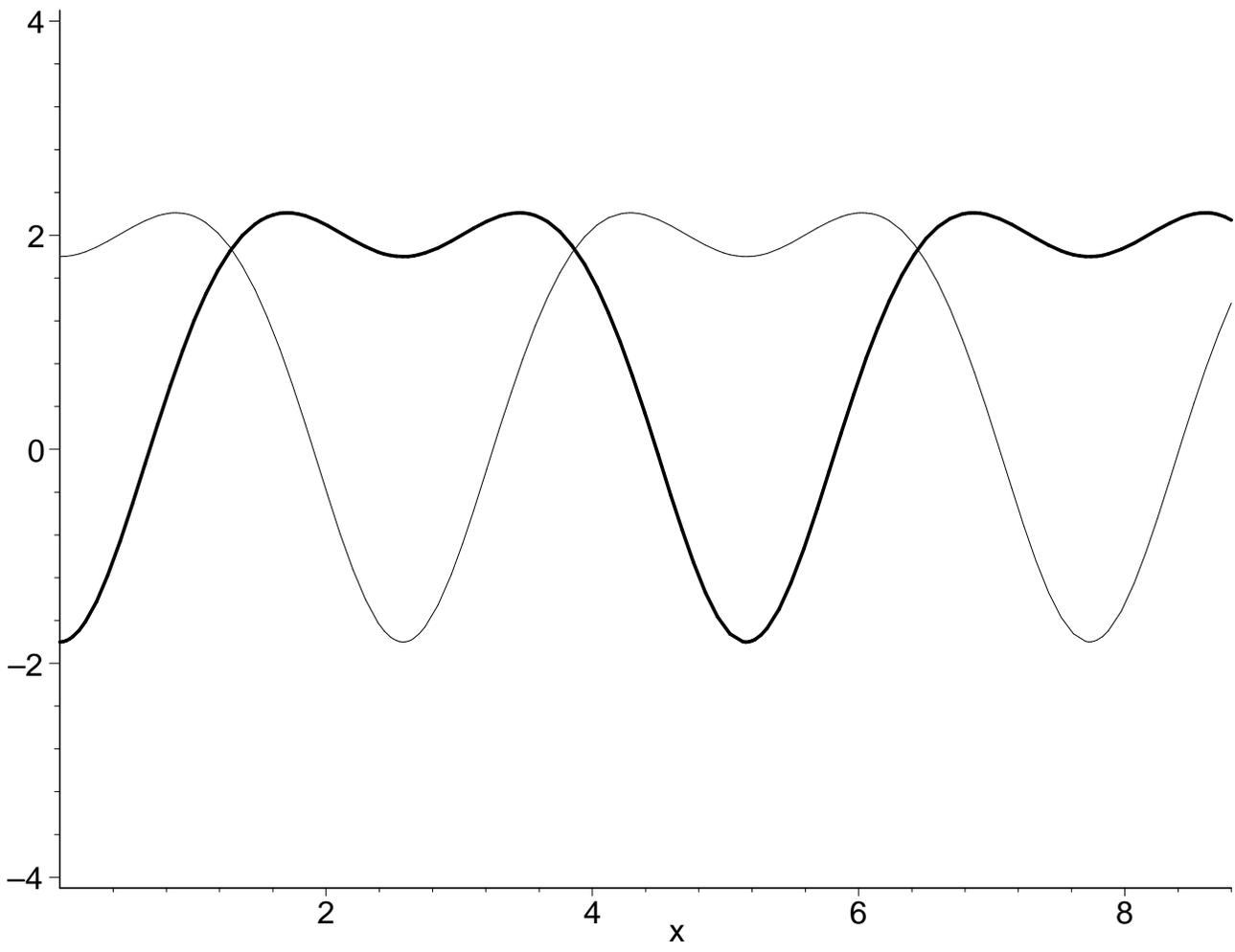
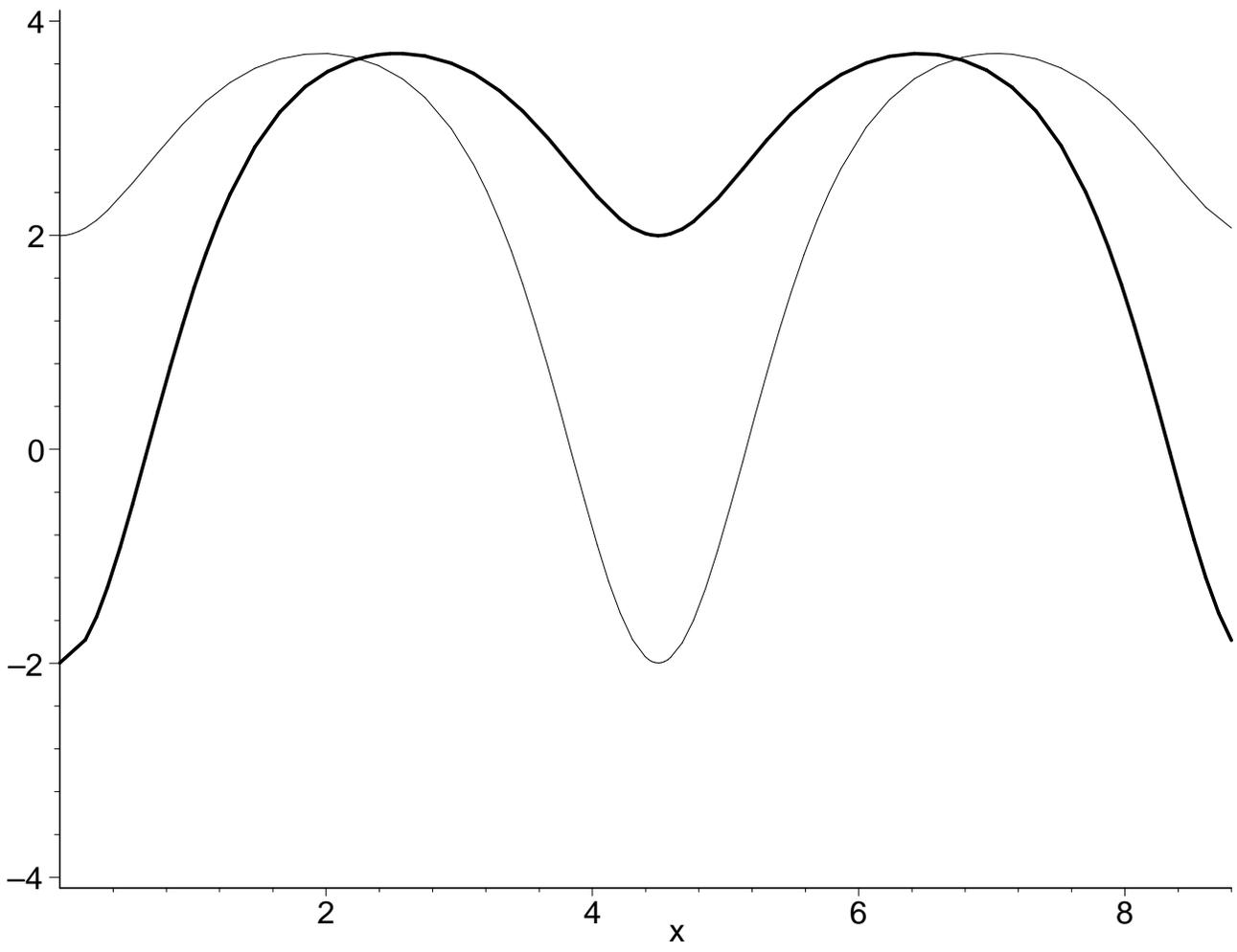
Fig. 6: The (6,2) associated Lam\'{e} potential $V_-(x)$ [thick line] and its supersymmetric partner potential 
$V_{+}(x)$ [thin line] 
for three choices of $m$ (a) 0.5 (b) 0.9 (c) 0.998. \sss

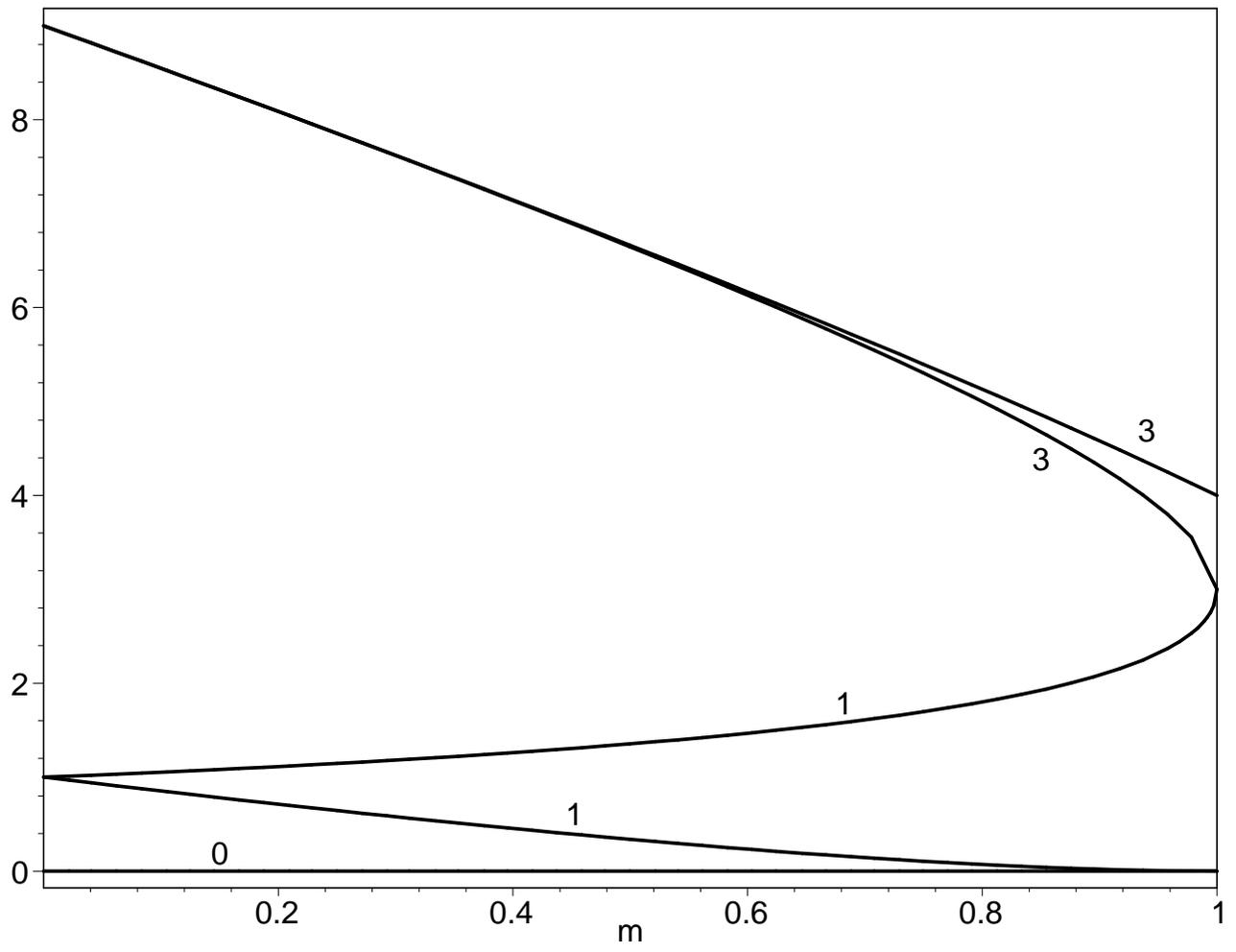
Fig. 7: Band edge energies for the associated Lam\'{e} potential (6,2) as a function of the elliptic modulus parameter $m$. This figure corresponds to Table 4. \sss

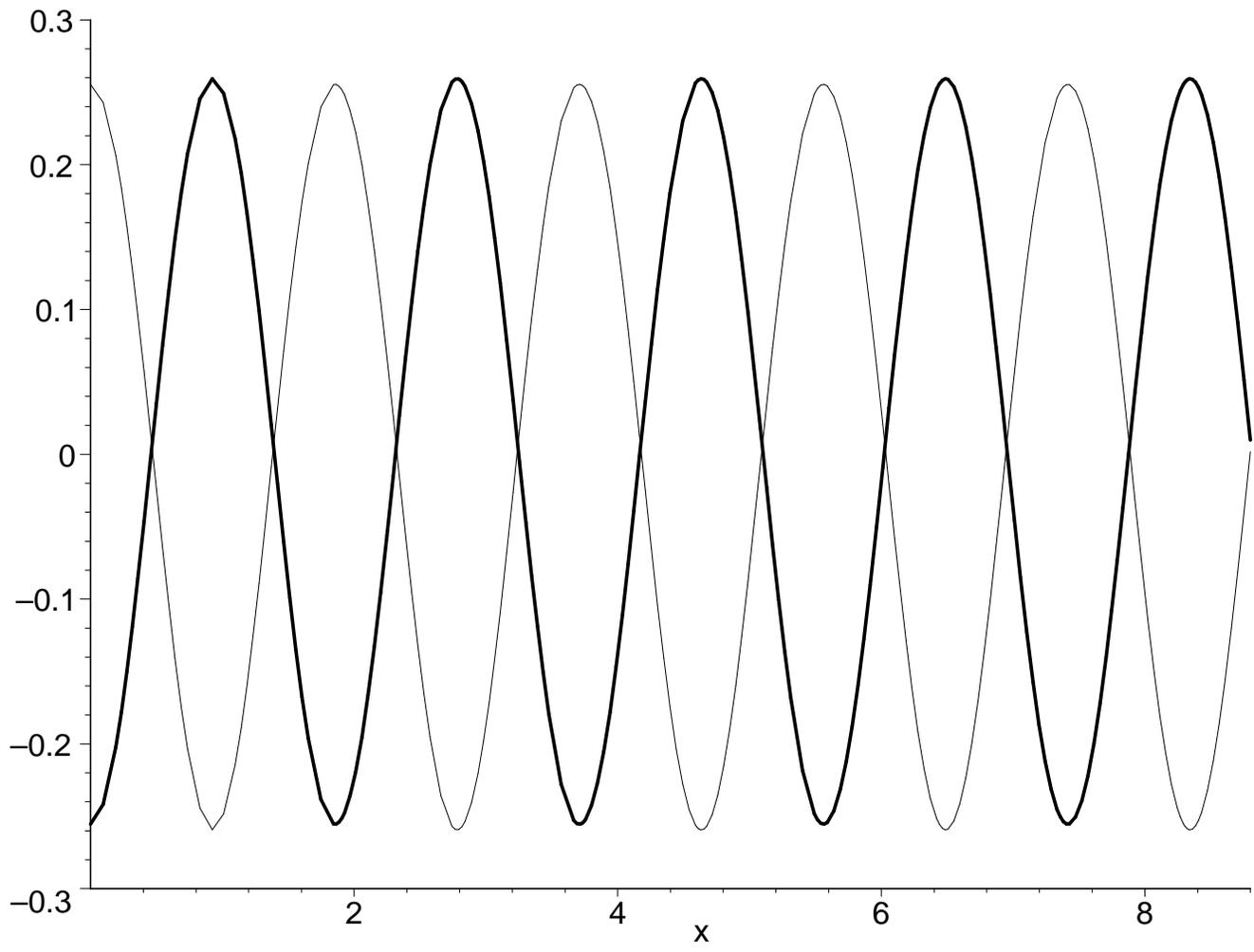
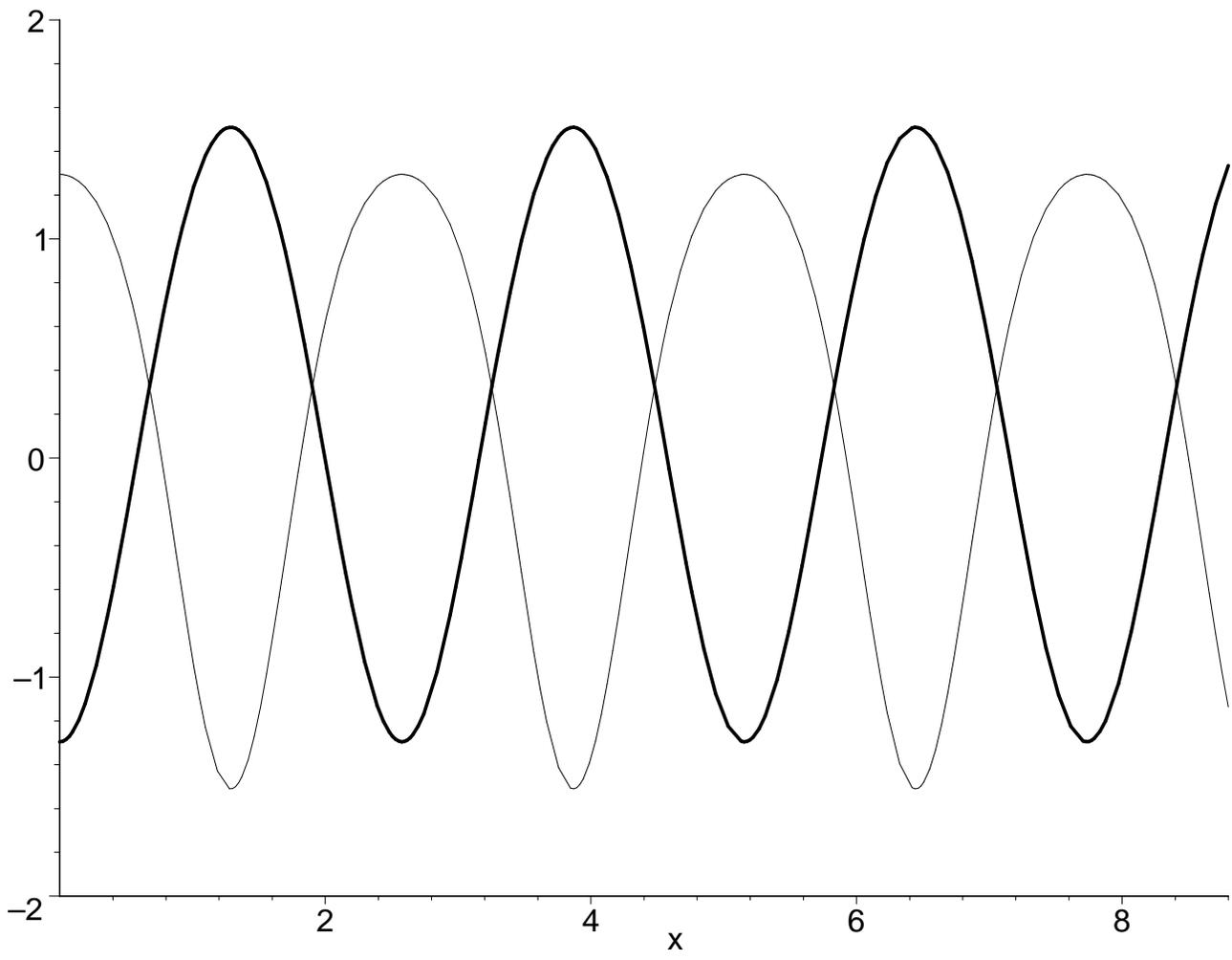
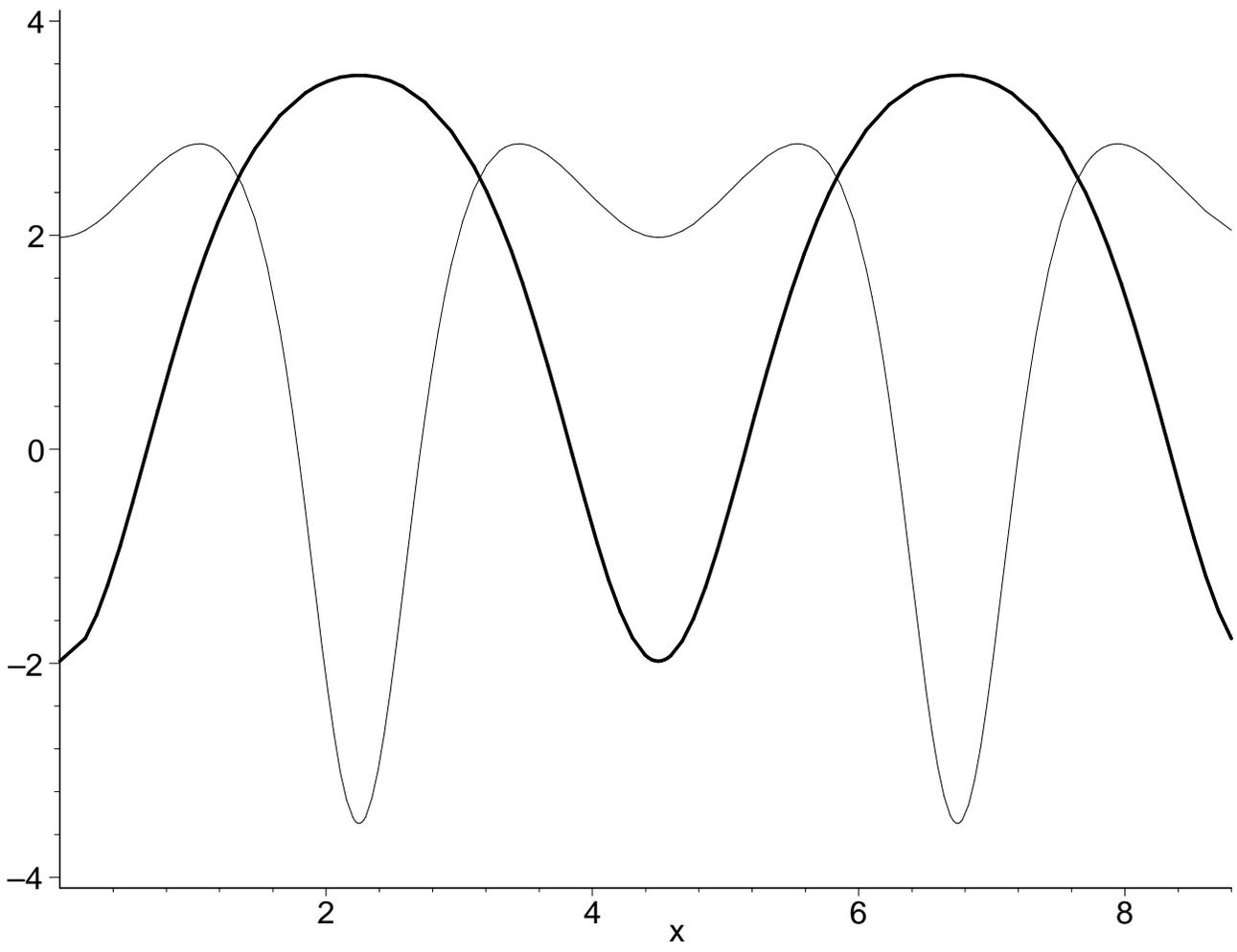
Fig. 8: The (6,6) associated Lam\'{e} potential $V_-(x)$ [thick line] as given by eq. (\ref{46}) and its supersymmetric partner potential $V_{+} (x)$ [thin line] 
as given by eq. (\ref{70}) for three choices of $m$ (a) 0.5 
(b) 0.9 (c) 0.998. \sss

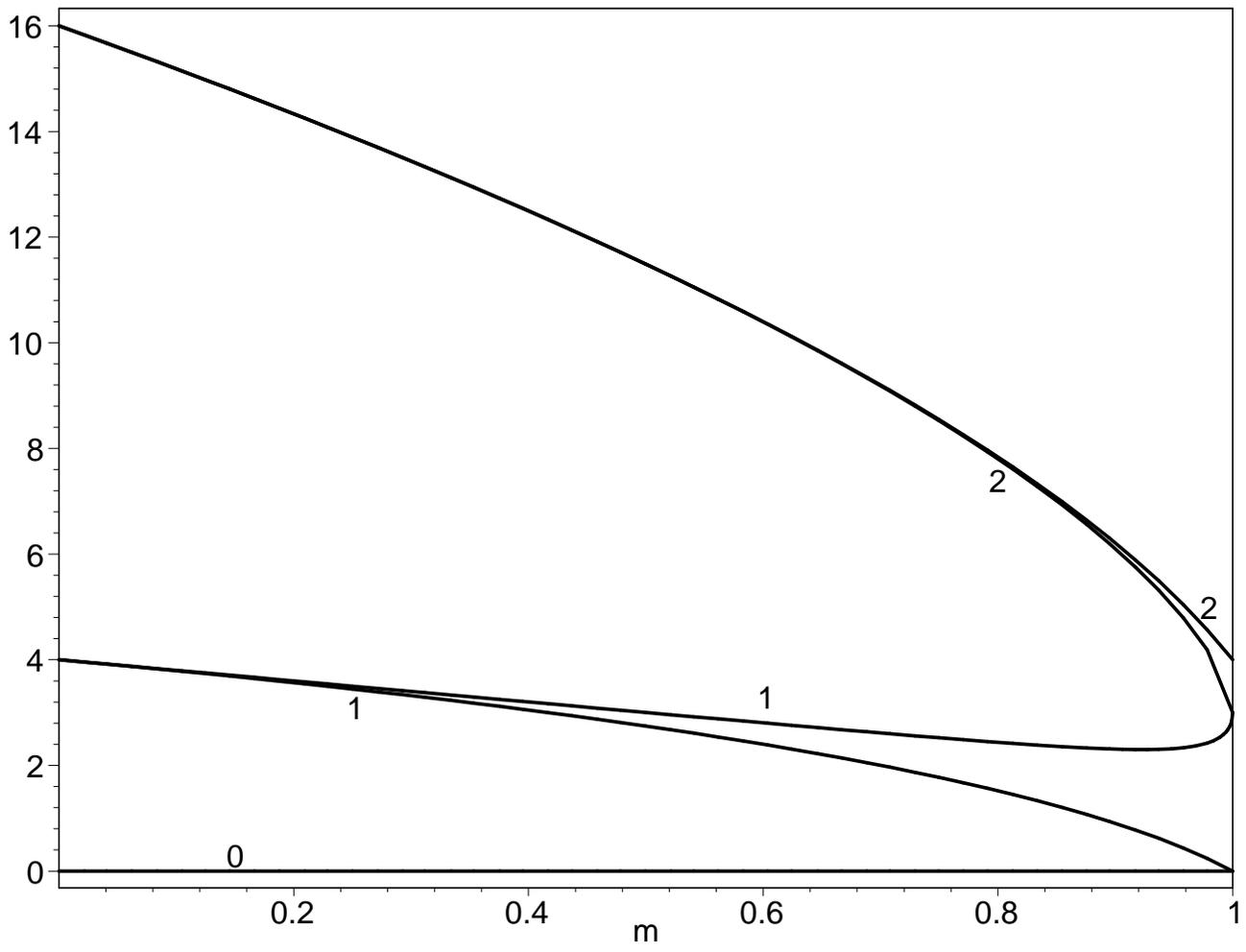
Fig. 9: Band edge energies for the associated Lam\'{e} potential (6,6) as a function of the elliptic modulus parameter $m$. This figure corresponds to Table 6. \sss

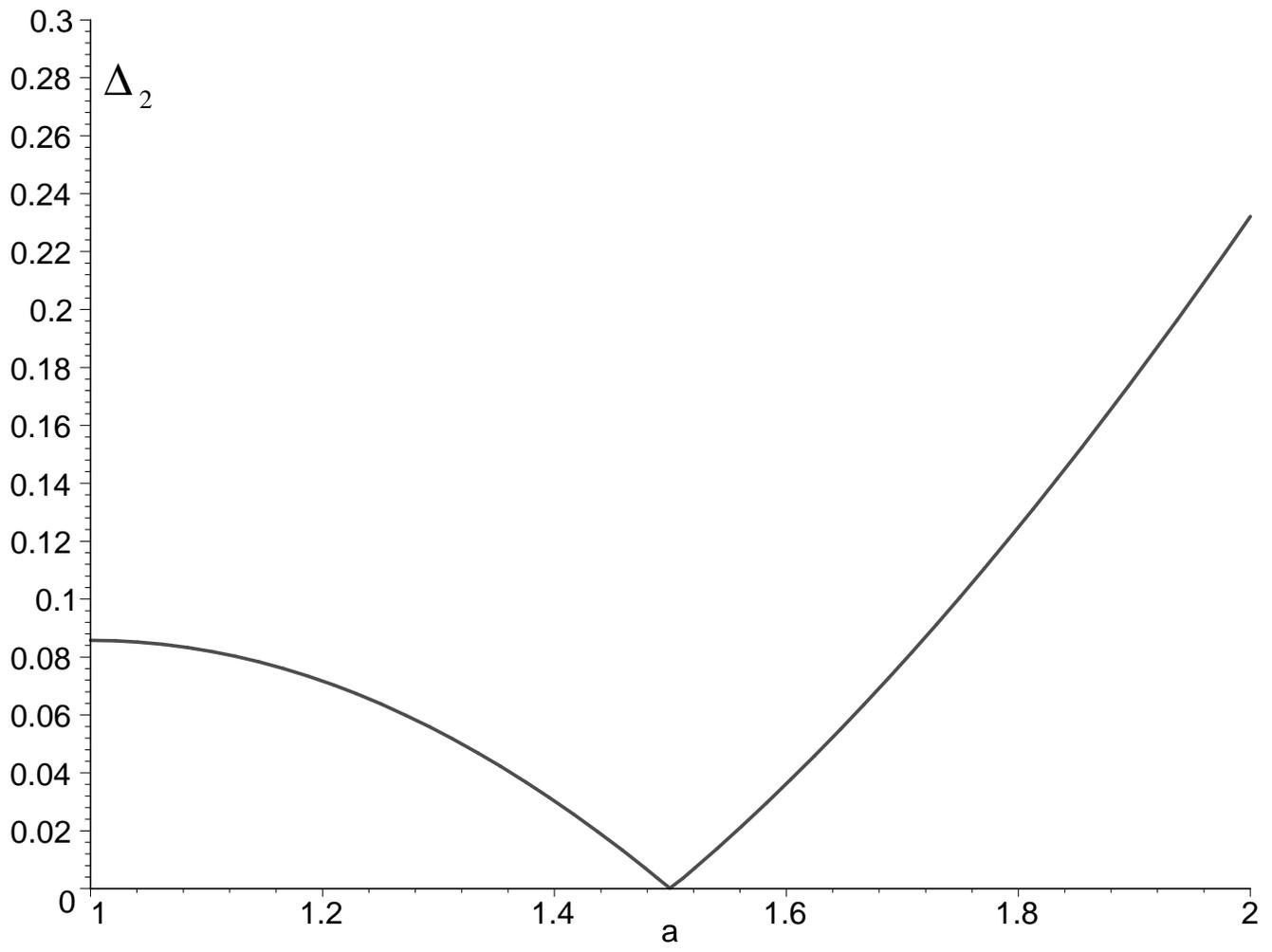
Fig. 10: Energy gap $\Delta_2 \equiv |E_4 - E_3|$ as one moves along the parabola of solvability $P2$ corresponding to $q = (a-2)(a-3)$ and $p = a(a+1)$. \sss

\newpage
\oddsidemargin      -0.5in
\centerline {\bf {Table 1: Energy Eigenstates for $V_{\pm}$ Corresponding to $a=2$}}
\bigskip
\begin{tabular}{cccccc}
\hline
 $E$ & $\psi^{(-)}$ & $[B-3m~{\rm sn}^2 (x)]\psi^{(+)}$ & 
${\rm Period}$ & ${\rm Nodes}$\\
\hline
 $0$ & $m + 1 +\delta-3m {\rm sn}^2 (x)$  & $1$  & $2K$ & $0$\\
 $ 2\delta-1-m$ & ${\rm cn}(x) {\rm dn}(x)$
& ${\rm sn}(x)[6m-(m+1)B+m {\rm sn}^2 (x) (2B-3-3m)]$ & $4K$ & $1$\\
 $ 2\delta-1+2m$ & ${\rm sn}(x) {\rm dn}(x)$
& $ {\rm cn}(x) [B+m {\rm sn}^2 (x) (3-2B)]$   & $4K$ & $1$\\
 $ 2\delta+2-m$ & ${\rm sn}(x) {\rm cn}(x)$
& ${\rm dn}(x) [B+{\rm sn}^2 (x) (3m-2B)]$   & $2K$ & $2$\\
 $4\delta$ & $m + 1 -\delta-3m {\rm sn}^2 (x)$ & ${\rm sn}(x)
{\rm cn}(x){\rm dn}(x)$   & $2K$ & $2$\\
\hline
\end{tabular}
\bigskip

\sss\sss\sss\sss
\oddsidemargin      -0.7in
\centerline {\bf {Table 2: Energy Eigenstates for $V_{\pm}$ Corresponding to $a=3$}}
\bigskip
\begin{tabular}{cccccc}
\hline
 $E$ & $\psi^{(-)}$ & $\psi_0^{-} \psi^{(+)}$ & 
${\rm Period}$ & ${\rm Nodes}$\\
\hline
 $0$ & $\dn (x) [1+2m +\delta_1 -5m {\rm sn}^2 (x)]$  & 1  & $2K$ & $0$\\
 $ 3-3m + 2\delta_1-2\delta_2$ & ${\rm cn}(x)[2+m+\delta_2 -5m \sn^2 (x)]$
& $10m(1-m+\delta_2 -\delta_1) {\rm sn}(x) \cn^2 (x) \dn^2 (x)$ 
& $4K$ & $1$ \\
 & & $-(1-m) {\sn (x) \psi^{-}_{0} \psi^{-} \over \cn (x) \dn (x)}$ 
 &  & \\
 $ 3+ 2\delta_1-2\delta_3$ & ${\rm sn}(x)[2+2m+\delta_3 -5m \sn^2 (x)]$
& $10m(1+\delta_3 -\delta_1) {\rm cn}(x) \sn^2 (x) \dn^2 (x)$
& $4K$ & $1$ \\
 & & $-(1-2m\sn^2 (x)) {\cn (x) \psi^{-}_{0} \psi^{-} \over \sn (x) \dn (x)}$ 
&  & \\
 $ 2-m+2\delta_{1}$ & ${\rm sn}(x) \cn (x) {\rm dn}(x)$
& $\dn^3 (x) [1+2m+\delta_1 +(m-2-2\delta_1) \sn^2 (x)]$ & $2K$ & $2$ \\
 4$\delta_1$ & $\dn (x) [1+2m -\delta_1 -5m {\rm sn}^2 (x)]$ 
& ${\rm sn}(x) \cn (x) \dn^3 (x)$ & 2K & $2$\\
 $ 3-3m + 2\delta_1+2\delta_2$ & ${\rm cn}(x)[2+m-\delta_2 -5m \sn^2 (x)]$
& $10m(1-m-\delta_2 -\delta_1) {\rm sn}(x) \cn^2 (x) \dn^2 (x)$ 
& $4K$ & $3$ \\
 & & $-(1-m) {\sn (x) \psi^{-}_{0} \psi^{-} \over \cn (x) \dn (x)}$ 
 &  & \\
 $ 3+ 2\delta_1+2\delta_3$ & ${\rm sn}(x)[2+2m-\delta_3 -5m \sn^2 (x)]$
& $10m(1-\delta_3 -\delta_1) {\rm cn}(x) \sn^2 (x) \dn^2 (x)$
& $4K$ & $3$ \\
 & & $-(1-2m\sn^2 (x)) {\cn (x) \psi^{-}_{0} \psi^{-} \over \sn (x) \dn (x)}$ 
&  & \\
\hline
\end{tabular}
\bigskip

\vskip 2.2 true cm
\newpage
\oddsidemargin      -0.1in
\centerline{ \bf {Table 3: Some Eigenstates for Various Associated 
Lam\'{e} Potentials}}
\bigskip
\begin{tabular}{ccccccc} 
\hline
 $q$ & $E$ & ${\rm dn}^{-a} (x) \psi $ & ${\rm Period}$ & ${\rm Nodes}$\\
\hline
 $a(a-1)$ & $ma^2$  & $1$ & $2K$ & $0$\\
 $(a-1)(a-2)$ & $1+m(a-1)^2$ & ${{\rm cn}(x)\over {\rm dn}(x)}$ & $4K$ & $1$\\ 
 $(a-1)(a-2)$ & $1+ma^2$ 
& $ {{\rm sn}(x)\over {\rm dn}(x)}$ & $4K$ & $1$\\
 $(a-2)(a-3)$ & $2+m(a^2-2a+2)\pm 2\delta_4$ 
& ${[m(2a-1) {\rm sn}^2 (x)-1+m-ma
\pm\delta_4]\over {\rm dn}^2 (x)}$ & $2K$ & $2,0$\\
 $(a-2)(a-3)$ & $4+m(a-1)^2$ & 
${{\rm sn}(x) {\rm cn}(x)\over {\rm dn}^2 (x)}$ & $2K$ & $2$\\
 $(a-3)(a-4)$ & $5+m(a^2-4a+5)\pm 2\delta_5$ 
& ${{\rm cn}(x)[m(2a-1) {\rm sn}^2 (x)-2+2m-ma \pm \delta_5]
\over {\rm dn}^3 (x)}$ & $4K$ & $3,1$\\
 $(a-3)(a-4)$ & $5+m(a^2-2a+2)\pm 2\delta_6$ 
& ${{\rm sn}(x)[m(2a-1) {\rm sn}^2 (x)-2+m-ma\pm \delta_6]
\over {\rm dn}^3 (x)}$ & $4K$ & $3,1$\\
$(a-4)(a-5)$ & $10+m(a^2-4a+5)\pm2\delta_7$ & ${{\rm sn} (x) {\rm cn} (x) 
[m(2a-1){\rm sn}^2 (x)-3+2m-ma\pm\delta_7 ] \over {\rm dn}^4 (x)}$ & $2K$ & $4,2$\\
\hline
\end{tabular}
\bigskip

\vskip 2.2 true cm
\centerline {\bf {Table 4: Energy Eigenstates for the $(6,2)$ Potential}}
\bigskip
\begin{tabular}{cccccc}
\hline
 $E$ & $\psi^{(-)}$ & 
${\rm Period}$ & ${\rm Nodes}$\\
\hline
 $0$ & $\dn^2 (x)$  
 & $2K$ & $0$\\
 $ 5-3m-2\sqrt{4-3m}$ & ${\cn (x)\over \dn (x)} [3m{\rm sn}^2 (x) -2
-\sqrt{4-3m}]$
 & $4K$ & $1$\\
 $ 5-2m-2\sqrt{4-5m+m^2}$ & ${{\rm sn}(x) \over {\rm dn}(x)}
[3m{\rm sn}^2 (x)-2-m-\sqrt{4-5m+m^2}]$
 & $4K$ & $1$\\
 $ 5-2m+2\sqrt{4-5m+m^2}$ & ${{\rm sn}(x) \over {\rm dn}(x)}
[3m{\rm sn}^2 (x)-2-m+\sqrt{4-5m+m^2}]$
 & $4K$ & $3$\\
 $ 5-3m+2\sqrt{4-3m}$ & ${\cn (x)\over \dn (x)} [3m{\rm sn}^2 (x) -2
+\sqrt{4-3m}]$
 & $4K$ & $3$\\
\hline
\end{tabular}
\bigskip

\vskip 2.2 true cm
\centerline{ \bf{Table 5: Energy Eigenstates for the $(2,2)$ Potential}}
\bigskip
\begin{tabular}{cccccc} 
\hline
 $E$ & ${\rm dn} (x) \psi^{(-)}$ 
 & ${\rm Period}$ & ${\rm Nodes}$\\
\hline
 $0$ & ${\rm dn}^2 (x)+\sqrt{1-m}$ & $K$ & $0$\\
 $4\sqrt{1-m}$ & ${\rm dn}^2 (x)-\sqrt{1-m}$ 
 & $2K$ & $1$\\
 $ 2-m+2\sqrt{1-m}$ & ${\rm sn}(x) {\rm cn}(x)$ 
 & $2K$ & $1$\\
\hline
\end{tabular}
\newpage

\centerline {\bf {Table 6: Energy Eigenstates for the $(6,6)$ Potential}}
\bigskip
\begin{tabular}{cccccc}
\hline
 $E$ & ${\rm dn}^2 (x) \psi^{(-)}$ & 
${\rm Period}$ & ${\rm Nodes}$\\
\hline
 $0$ & $1 -(4-m-\delta_8) {\rm sn}^2 (x)+ (4-2m-\delta_8) {\rm sn}^4 (x)$  
 & $K$ & $0$\\
 $ -4+2m+2\delta_8$ & $1 - 2{\rm sn}^2 (x) + m {\rm sn}^4 (x)$
 & $2K$ & $1$\\
 $ 2-m-6\sqrt{1-m} +2\delta_8$ & ${\rm sn}(x) {\rm cn}(x) [1-(1-\sqrt{1-m})
{\rm sn}^2 (x)]$
 & $2K$ & $1$\\
 $ 2-m+6\sqrt{1-m} +2\delta_8 $ & ${\rm sn}(x) {\rm cn}(x) [1-(1+\sqrt{1-m})
{\rm sn}^2 (x)]$
 & $K$ & $2$\\
 $4\delta_8$ & $ 1 -(4-m+\delta_8) {\rm sn}^2 (x) + (4-2m +\delta_8) 
{\rm sn}^4 (x)$ 
 & $K$ & $2$\\
\hline
\end{tabular}
\bigskip

\vskip 2.2 true cm
\centerline {\bf{Table 7: Energy Eigenstates for the $(63/4,3/4)$ Potential}}
\bigskip
\begin{tabular}{cccccc}
\hline
 $E$ & ${\rm dn}^{1/2} (x)\psi^{(-)}$ & 
${\rm Period}$ & ${\rm Nodes}$\\
\hline
 $0$ & $[12m \sn^2 (x)-2-5m-\delta_9] {\rm dn}^2 (x)$ & $2K$ & $0$\\
 $2-m+\delta_9$ & ${\rm sn} (x) {\rm cn} (x) {\rm dn}^2 (x)$  & $2K$ & $2$\\
 $2\delta_9$ & $[12m {\rm sn}^2 (x)-2-5m+\delta_9] {\rm dn}^2 (x)$  & $2K$ & $2$\\
 $14-7m+\delta_9$ & ${\rm sn} (x) {\rm cn} (x) [1-2 {\rm sn}^2 (x)]$  & $2K$ & $4$\\
 $14-7m+\delta_9$ & $[1-8 {\rm sn}^2 (x) {\rm cn}^2 (x)]$  & $2K$ & $4$\\
\hline
\end{tabular}
\bigskip

\vskip 1.1 true cm

\begin{thebibliography}{99}
\bibitem{ff} Some isolated 
solutions of the Schr\"odinger 
equation for Lam\'{e} potentials with half-integer 
values of $a$ are also
known \cite{ar}, but they have certain peculiar features which require further study.
\end{thebibliography}
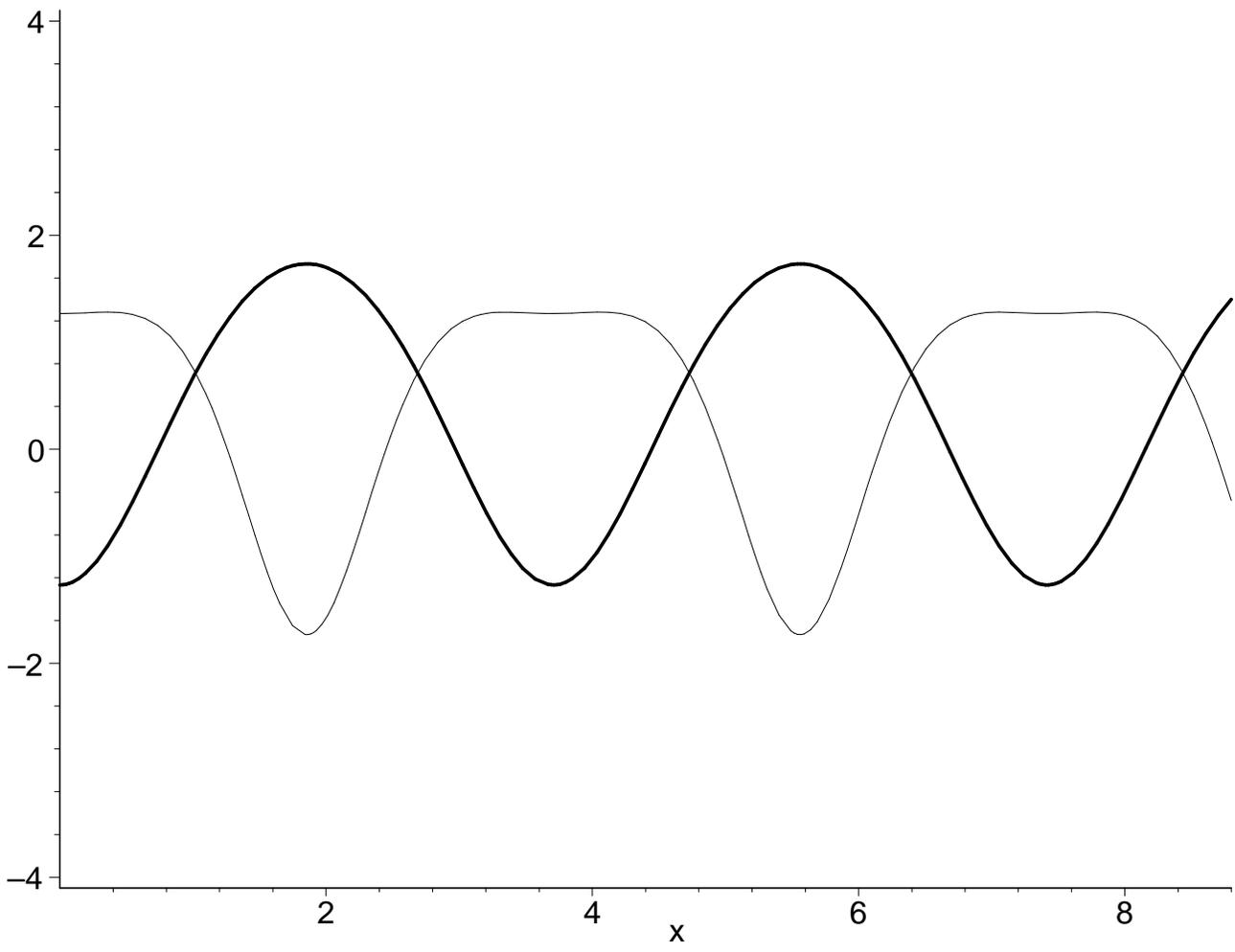
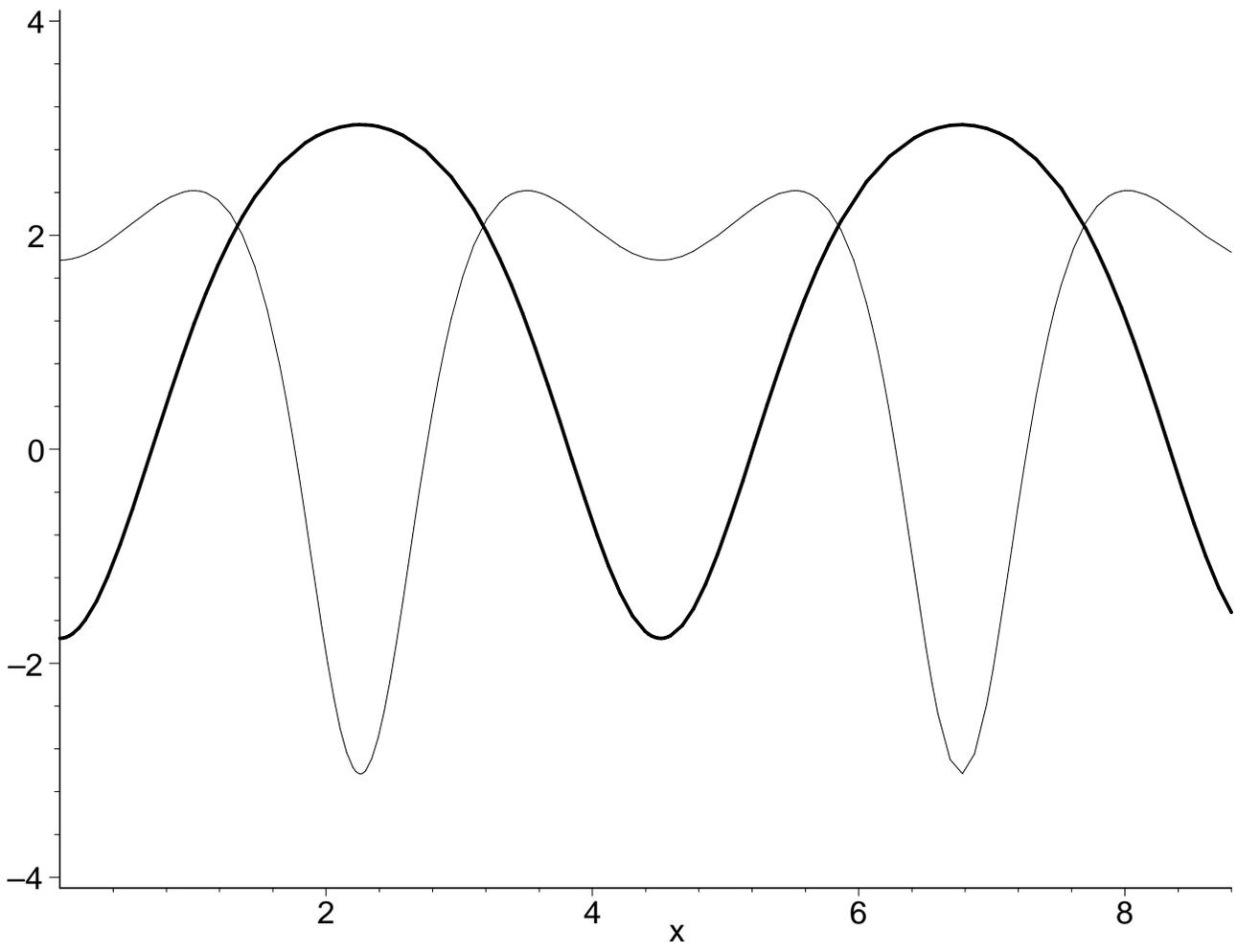
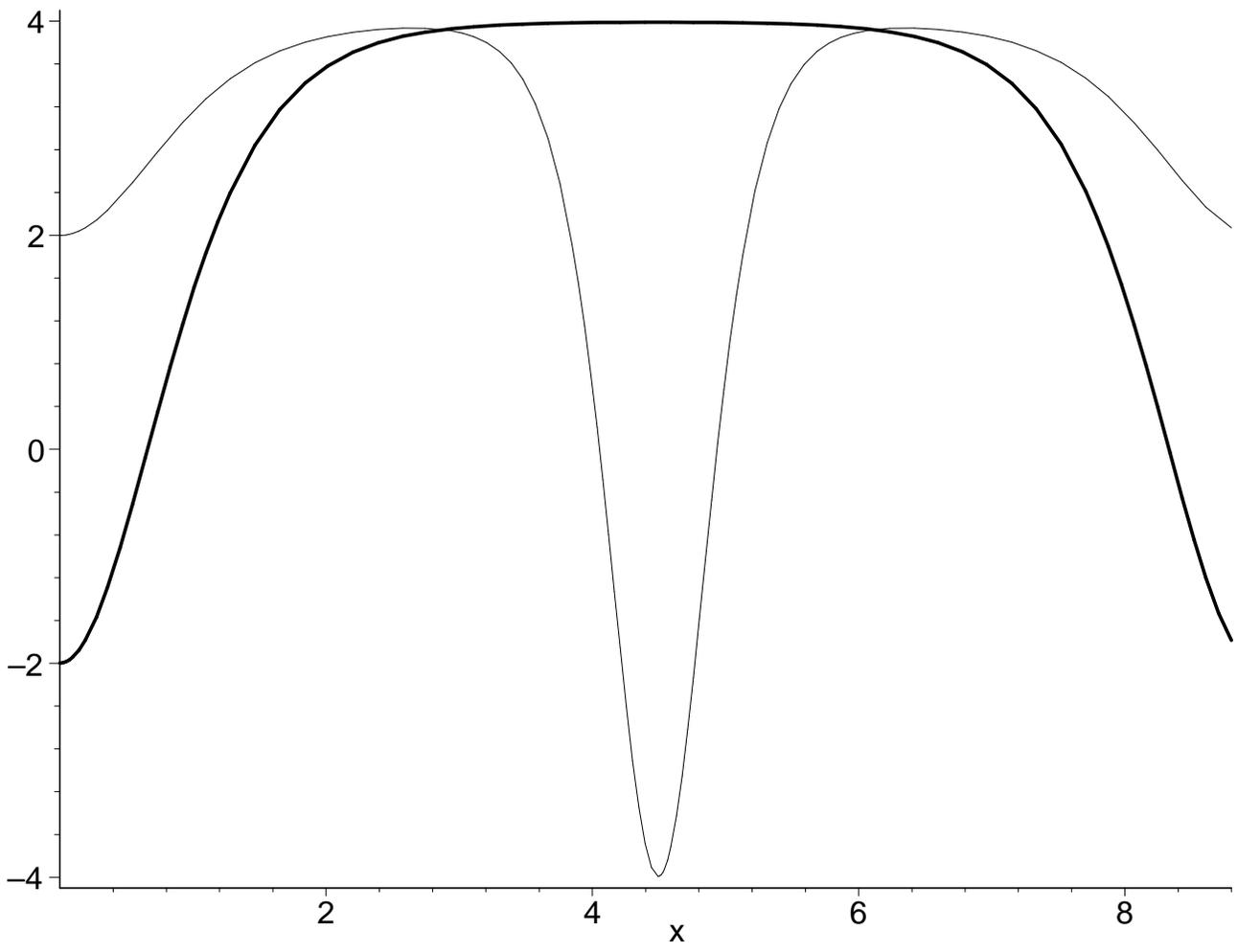
\end{document}